\documentclass[
,aps
,pra
,amsmath
,eqsecnum
,amssymb
,amsfonts
,twocolumn
,letterpaper
,10pt
,tightenlines
,superscriptaddress
,nofootinbib
,eqsecnum
,longbibliography
,notitlepage
,floatfix
]{revtex4-2}

\usepackage{nag}
\usepackage{graphicx}
\usepackage{subfigure}
\usepackage{amsthm}
\usepackage{tabulary}
\usepackage{threeparttable}
\usepackage{qcircuit}
\usepackage{mathtools}
\usepackage{bm}
\usepackage[table,xcdraw]{xcolor}
\usepackage{datetime}
\usepackage{stackrel}
\usepackage{stmaryrd}
\usepackage{dsfont}
\usepackage{qcircuit}
\usepackage[english]{babel}
\usepackage{units}

\usepackage{tikz}
\usetikzlibrary{backgrounds,decorations.pathreplacing}

\usepackage{chngcntr}

\usepackage[normalem]{ulem}
\usepackage{slashed}
\usepackage{soul}

\usepackage{xcolor}
\usepackage[
    colorlinks=true,
    urlcolor=blue,
    linkcolor=blue,
    citecolor=blue,
    filecolor=blue,
]{hyperref}
\usepackage{amssymb}
\usepackage{amsmath}
\usepackage{amsthm}

\usepackage[OT2,T1]{fontenc}
\DeclareSymbolFont{cyrletters}{OT2}{wncyr}{m}{n}
\DeclareMathSymbol{\Sha}{\mathalpha}{cyrletters}{"58}

\usepackage{makeidx}
\makeindex

\usepackage{soul}
\usepackage{xargs}
\newcommandx{\cmnote}[2][1=]{\linespread{1.0}\todo[linecolor=red,backgroundcolor=red!25,bordercolor=red,#1]{#2}}

\let\underline\ul

\allowdisplaybreaks[4]

\DeclareMathOperator{\Tr}{Tr}

\renewcommand{\Re}{\operatorname{Re}}
\renewcommand{\Im}{\operatorname{Im}}

\let\originalleft\left
\let\originalright\right
\renewcommand{\left}{\mathopen{}\mathclose\bgroup\originalleft}
\renewcommand{\right}{\aftergroup\egroup\originalright}

 \index{} \index{} \index{} \index{} \index{} \index{} \index{} \index{}%
\makeatletter

\newcommand{\ringplus}{\mathbin{\text{\@ringplus}}}

\newcommand{\@ringplus}{%
  \ooalign{\hidewidth\raise1.3ex\hbox{\tiny$\circ$}\hidewidth\cr$\m@th+$\cr}%
}

\newcommand{\ringminus}{\mathbin{\text{\@ringminus}}}

\newcommand{\@ringminus}{%
  \ooalign{\hidewidth\raise0.9ex\hbox{\tiny$\circ$}\hidewidth\cr$\m@th-$\cr}%
}
\makeatother
 \index{} \index{} \index{} \index{} \index{} \index{} \index{} \index{}%

\newcommand{\tp}[0]{\mathrm{T}}

\DeclareFontFamily{U}{wncy}{}
\DeclareFontShape{U}{wncy}{m}{n}{<->wncyr10}{}
\DeclareSymbolFont{mcy}{U}{wncy}{m}{n}
\DeclareMathSymbol{\Sh}{\mathord}{mcy}{"58}

\newcommand{\negspace}{\!}

\newcommand{\lsub}[2]{{\protect\vphantom{#1}}_{#2} \negspace {#1}}
\newcommand{\rsub}[2]{{#1} \negspace {\protect\vphantom{#1}}_{#2}}
\newcommand{\lrsub}[3]{{\protect\vphantom{#1}}_{#2} \negspace {#1} \negspace {\protect\vphantom{#1}}_{#3}}

\newcommand{\ketsub}[2]{\rsub {\ket{#1}} {#2}}
\newcommand{\brasub}[2]{\lsub {\bra{#1}} {#2}}

\newcommand{\pbra}[1]{\brasub{#1} p}

\newcommand{\inprod}[2]{\left\langle {#1} | {#2} \right\rangle}

\newcommand{\inprodsubsub}[4]{\lrsub {\inprod{#1}{#2}} {#3} {#4}}

\newcommand{\outprod}[2]{\ket {#1}\!\bra {#2}}

\newcommand{\outprodsubsub}[4]{\ketsub {#1}{#3} \brasub{#2}{#4}}

\newcommand{\avg}[1]{\left\langle {#1} \right\rangle}

\newcommand{\reals}[0]{\mathbb{R}}
\newcommand{\complex}[0]{\mathbb{C}}

\newcommand{\integers}[0]{\mathbb{Z}}
\newcommand{\rationals}[0]{\mathbb{Q}}

\newcommand{\op}[1]{\hat{#1}}

\newcommand{\opvec}[1]{\op{\vec{#1}}}

\newcommand{\mat}[1]{\bm{\mathrm{#1}}}

\renewcommand{\vec}[1]{\bm{#1}}

 \index{} \index{} \index{} \index{} \index{} \index{} \index{} \index{}%
 \index{} \index{} \index{} \index{} \index{} \index{} \index{} \index{}%

 \index{} \index{} \index{} \index{} \index{} \index{} \index{} \index{}%
 \index{} \index{} \index{} \index{} \index{} \index{} \index{} \index{}%
\usepackage{graphicx}

\usepackage{graphicx}
\usepackage{multirow}
\usepackage{hhline}

\xyoption{color}
\xyoption{line}

\newcommandx*\bsbal[3][1=black, 3=->]{\ar @[#1]@{#3} [#2,0] \qw}

\newcommandx*\varbs[4][1=black, 3=\theta, 4=->]{\ar @[#1]@{#4}^{#3} [#2,0] \qw}
\newcommandx*\varbsleft[5][1=black, 3=\theta', 4=->]{\ar @[#1]@{#4}^{#3}_{#5} [#2,0] \qw}

\newcommandx*\ctrlg[2]{\control \ar @{-}^{#1} [#2,0] \qw}
\newcommandx*\ctrlog[2]{\controlo \ar @{-}^{#1} [#2,0] \qw}

\newcommand{\nogate}[1]{*+<.6em>{#1} \POS ="i","i"+UR;"i"+UL **[white]\dir{-};"i"+DL **[white]\dir{-};"i"+DR **[white]\dir{-};"i"+UR **[white]\dir{-},"i" }

\makeatletter
 \newcommand{\xmapsfrom}[2][]{%
    \ext@arrow3095\leftarrowfill@{#1}{#2}\mapsfromchar
}
\makeatother

\usepackage{mathtools}

\DeclarePairedDelimiter{\bra}{\langle}{\rvert}%
\DeclarePairedDelimiter{\ket}{\lvert}{\rangle}%
\DeclarePairedDelimiterX\innerp[2]{\langle}{\rangle}{#1\delimsize\vert\mathopen{}#2}%
\DeclarePairedDelimiterX\braket[2]{\langle}{\rangle}{#1\delimsize\vert\mathopen{}#2}%
\DeclarePairedDelimiterX\braketOP[3]{\langle}{\rangle}{#1\,\delimsize\vert\,\mathopen{}#2\,\delimsize\vert\,\mathopen{}#3}%
\DeclarePairedDelimiterX\ketbra[2]{\lvert}{\rvert}{#1\delimsize\rangle\!\delimsize\langle#2}%
\DeclarePairedDelimiterX\outerp[2]{\lvert}{\rvert}{#1\delimsize\rangle\!\delimsize\langle#2}%
\DeclarePairedDelimiterX\projector[1]{\lvert}{\rvert}{#1\delimsize\rangle\!\delimsize\langle#1}%
\usepackage{float}

\usepackage{comment}
\usepackage{placeins}
\usepackage{flushend}
\usepackage{leftidx}
\usepackage{etoolbox}
\usepackage{bm}
\usepackage{makecell}
\let\tinymatrix\smallmatrix

\patchcmd{\tinymatrix}{\scriptstyle}{\scriptscriptstyle}{}{}
\patchcmd{\tinymatrix}{\scriptstyle}{\scriptscriptstyle}{}{}
\patchcmd{\tinymatrix}{\vcenter}{\vtop}{}{}
\patchcmd{\tinymatrix}{\bgroup}{\bgroup\scriptsize}{}{}

\DeclareMathOperator*{\sumint}{%
\mathchoice%
{\ooalign{$\displaystyle\sum$\cr\hidewidth$\displaystyle\int$\hidewidth\cr}}
  {\ooalign{\raisebox{.14\height}{\scalebox{.7}{$\textstyle\sum$}}\cr\hidewidth$\textstyle\int$\hidewidth\cr}}
  {\ooalign{\raisebox{.2\height}{\scalebox{.6}{$\scriptstyle\sum$}}\cr$\scriptstyle\int$\cr}}
  {\ooalign{\raisebox{.2\height}{\scalebox{.6}{$\scriptstyle\sum$}}\cr$\scriptstyle\int$\cr}}
}

\newcommand{\projGKP}{\bar{\Pi}}
\newcommand{\pauliGKP}{\bar{\sigma}}

\newcommand{\qubitKraus}{\bar{K}}
\newcommand{\qubitChannel}{\bar{\mathcal{E}}}
\newcommand{\krausEC}{\op{K}_\text{EC}}
\newcommand{\krausloss}{\op{L}}

\newcommand{\paulivec}{\vec{a}}

\newcommand{\het}{\text{Het}}
\newcommand{\photon}{\text{P}}
\newcommand{\qad}{\text{Hom}}
\newcommand{\qadrot}{\text{Hom}_\phi}

\newcommand{\krauslossphoton}{\krausloss^\photon}
\newcommand{\krauslosshet}{\krausloss^\het}
\newcommand{\krauslossqad}{\krausloss^\qad}
\newcommand{\krauslossqadrot}{\krausloss^{\qadrot}}
\newcommand{\qubitKrausphoton}{\qubitKraus^\photon}
\newcommand{\qubitKraushet}{\qubitKraus^\het}

\renewcommand{\selectlanguage}[1]{}
\begin{document}
\title{Logical channel for heralded and pure loss with the Gottesman-Kitaev-Preskill code}

\author{Tom B. Harris}
\email{tombharris059@gmail.com}
\affiliation{Centre for Quantum Computation and Communication Technology, School of Science, RMIT University, Melbourne, VIC 3000, Australia}
\author{Takaya Matsuura}
\affiliation{Centre for Quantum Computation and Communication Technology, School of Science, RMIT University, Melbourne, VIC 3000, Australia}
\affiliation{RIKEN Center for Quantum Computing (RQC), Hirosawa 2-1, Wako, Saitama 351-0198, Japan}
\author{Ben Q. Baragiola}
\affiliation{Centre for Quantum Computation and Communication Technology, School of Science, RMIT University, Melbourne, VIC 3000, Australia}
\author{Nicolas C. Menicucci}
\affiliation{Centre for Quantum Computation and Communication Technology, School of Science, RMIT University, Melbourne, VIC 3000, Australia}

\setcounter{page}{1}
\pagenumbering{arabic}
\begin{abstract}
Photon loss is the dominant source of noise in optical quantum systems. The Gottesman-Kitaev-Preskill (GKP) bosonic code provides significant protection; however, even low levels of loss can generate uncorrectable errors that another concatenated code must handle.
In this work, we characterize these errors by deriving analytic expressions for the logical channel that arises from pure loss acting on approximate GKP qubits. Unlike random displacement noise, we find that the loss-induced logical channel is not a stochastic Pauli channel.
We also provide analytic expressions for the logical channel for ``heralded loss,'' when the light scattered out of the signal mode is measured either by photon number counting---\emph{i.e.} photon subtraction---or heterodyne detection. These offer a pathway to intentionally inducing non-Pauli channels for, \emph{e.g.}, magic-state production.

\end{abstract}
\maketitle

\section{Introduction}
Continuous-variable (CV) quantum computing has emerged as a compelling approach for encoding and processing quantum information \cite{braunstein_quantum_2005,weedbrook_gaussian_2012,heeres_implementing_2017,terhal_towards_2020,gertler_protecting_2021,Xanadu2025scaling}. Bosonic modes are ubiquitous in nature and naturally support infinite-dimensional Hilbert spaces that can be leveraged for error correction using bosonic codes \cite{cochrane_macroscopically_1999, gottesman_encoding_2001, albert_bosonic_2022,michael_new_2016, bergmann_quantum_2016, fukui_analog_2017, noh_encoding_2020}.
Among bosonic codes, the Gottesman-Kitaev-Preskill (GKP) code stands out for its ease of use:
Clifford gates and Pauli measurements require only Gaussian operations, and universality is achieved by supplementing with the trivial vacuum state~\cite{baragiola_all-gaussian_2019}. These advantages are natural in optical systems, 
and recent experimental advances indicate that optical GKP state generation may soon be within reach~\cite{konno_logical_2024}.
Meanwhile, high-quality GKP states are routinely produced and manipulated in trapped ions and microwave cavities \cite{fluhmann_encoding_2019,campagne-ibarcq_quantum_2020,gao_entanglement_2019,Matsos2024GKPprep}, where strong coupling to a qubit provides a lever for universal control. 

Although GKP codes were originally designed to correct small displacements in phase space \cite{gottesman_encoding_2001}, it was discovered that they also reign supreme against loss~\cite{albert_performance_2018}, outperforming other single-mode bosonic codes such as cat \cite{li_cat_2017,hastrup_all-optical_2022} and binomial codes \cite{michael_new_2016} when the code quality is high enough. 
This is critical because photon loss---arising from imperfect transmission through fibers, beam splitters, and other components---is the dominant noise source in optical systems \cite{knill_scheme_2001}. 
And unlike random displacements, which add noise incoherently, 
loss contracts a quantum state in phase space toward vacuum, fundamentally altering its structure~\cite{eisert_gaussian_2005, hastrup_analysis_2023}.

How, exactly, does loss affect quantum information encoded in the GKP code? The answer in general for any quantum error-correcting code can be quantified by finding the \emph{logical channel} given by encoding $\mathcal{E}$, applying physical noise, $\mathcal{N}$, and then performing decoding and recovery, $\mathcal{R}$: $\mathcal{E}_\text{L} = \mathcal{R} \circ \mathcal{N} \circ \mathcal{E}$~\cite{Doherty2002logicalchannel}. 
In this work, we analyze the GKP logical channel induced by damping and pure loss\, where damping refers to the operator $e^{-\beta\op{n}}$, which is commonly employed to describe finite-energy, approximate GKP states \cite{menicucci_fault-tolerant_2014,matsuura_equivalence_2020}. We derive analytic expressions for the logical maps post-selected on syndrome outcomes, using ideal GKP error correction \cite{gottesman_encoding_2001} as the recovery.  
These conditional maps can be numerically integrated over syndromes to characterize the average performance of GKP error correction under damping and loss. 
Our analytical framework also allows us to tackle ``heralded loss,'' where we get knowledge about the lost photons by performing measurements of the mode into which they are scattered.
Depending on the measurement basis and the outcome, GKP error correction can produce a highly non-Pauli logical channel that could be useful for generating GKP magic states. 
\section{Preliminaries}\label{sec:2.9Preliminaries}

\subsection{The GKP code}\label{sec:2.9GKP}

We briefly review the GKP code; more details can be found in the original paper~\cite{gottesman_encoding_2001} and from the vast amount of work that has followed~\cite{brady2024GKPreview}. The computational basis states of the square-lattice GKP code can be expressed as periodic superpositions of position $\ket{\cdot}_q$ or momentum eigenstates $\ket{\cdot}_p$,
    \begin{align} 
        \ket{\bar{a}} &\propto \sum_{n\in\mathbb{Z}} \ket{(2n+a)\sqrt{\pi}}_q\propto\sum_{n\in\mathbb{Z}} (-1)^{an}\ket{n\sqrt{\pi}}_p, \label{eq:Ideal_GKP}
    \end{align}
where $a\in\{0,1\}$.%
\footnote{\label{foot:qp}The position and momentum operators are defined as ${\op q = (\op a + \op a^\dag) / \sqrt 2}$ and ${\op p = -i (\op a - \op a^\dag) / \sqrt 2}$, respectively, so that $[\op q, \op p] = i$ (with $\hbar = 1$), and the vacuum variance is $\avg {\op q^2}_{\text{vac}} = \avg {\op p^2}_{\text{vac}} = \tfrac 1 2$.}
The dual-basis states are defined, as usual, as $\ket{\bar{\pm}} = \frac{1}{\sqrt{2}}(\ket{\bar{0}}\pm\ket{\bar{1}}).$ The standard Glauber displacement operator~\cite{walls_quantum_2008}, $\op D(\alpha) = e^{\alpha \op a^\dag - \alpha^* \op a}$, will be written in this work as taking a vector input~${\vec x = (\Re \alpha, \Im \alpha)^\tp}$ that characterizes the displacement amplitude:%
    \begin{align}
        \op D(\alpha) \mapsto \op{D}(\vec{x}) = e^{-i \sqrt{2} \vec{x}^\tp \mat{\Omega} \opvec{x}} = e^{-i \sqrt{2} (x_1 \op{p} - x_2 \op{q}) },\label{eq:DisplacementOp}
    \end{align}
where $\opvec x = (\op q, \op p)^\tp$, and $\mat \Omega$ is the matrix representing the symplectic form,
    \begin{align} \label{eq:symplectic-form}
        \mat{\Omega} \coloneqq \begin{bmatrix} 0 & 1 \\ -1 & 0 \end{bmatrix}.
    \end{align}
Note that the factor of~$\sqrt{2}$ comes from the definition of $\hat{q}$ and $\hat{p}$ (see footnote~\ref{foot:qp}).
The GKP logical Pauli gates are implemented by restricting the elements of $\vec{x}$ to odd integer multiples of $\sqrt{\pi/2}$. With this in mind, we define \emph{Pauli shift operators}
    \begin{align} \label{eq:Paulishift}
        \op{D}\left( \paulivec \sqrt{\frac{\pi}{2}}\right) = e^{-i \sqrt{\pi} (a_1 \op{p} - a_2 \op{q})},
    \end{align}
where $\paulivec\in \mathbb{Z}^2$ determines which Pauli is implemented within the GKP code space. A minimal representative set of such shifts covering all the Paulis is given by
    \begin{align} \label{eq:logical-shifts}
        \op{I}_L: \,\, &\paulivec_0 = (0,0), \quad \quad  
        \op{X}_L: \,\,  \paulivec_1 = (1,0), \\
        \op{Y}_L: \,\, &\paulivec_2 = (1,1), \quad \quad
        \op{Z}_L: \,\,  \paulivec_3 = (0,1).
    \end{align}
The stabilizers for the square-lattice GKP code, 
\begin{align}
    \op{S}_X &\coloneqq \op{D} \big( \paulivec_1 \sqrt{2\pi} \big) = e^{-i2\sqrt{\pi} \op{p}}, \\
    \op{S}_Z &\coloneqq \op{D} \big( \paulivec_3 \sqrt{2\pi} \big) = e^{i2\sqrt{\pi} \op{q}} ,
\end{align}
respectively implement shifts by 2$\sqrt{\pi}$ in position and momentum, and as such, they commute, $[\op{S}_X, \op{S}_Z] = 0$.

The projector onto the GKP subspace is formed using the computational states in Eq.~\eqref{eq:Ideal_GKP}, $\projGKP = \outprod{\bar{0}}{\bar{0}} + \outprod{\bar{1}}{\bar{1}}$, or as a sum over all powers of the stabilizers\footnote{The factor of $1/\sqrt{\pi}$ comes from the chosen normalization on the computational basis states.}  
    \begin{align} 
        \projGKP 
        = \frac{1}{\sqrt{\pi}} \sum_{n_1,n_2\in\mathbb{Z}^2} \big( \op{S}_X \big)^{n_1} \big( \op{S}_Z \big)^{n_2} 
    = \frac{1}{\sqrt{\pi}}\sum_{\vec{n}\in\mathbb{Z}^2} \op{D} \big( \vec{n}\sqrt{2\pi} \big)
    .\label{eq:GKPprojector}
    \end{align}
Projection of the Pauli shift operators into the GKP subspace yields the \emph{GKP subspace Paulis}~\cite{baragiola_all-gaussian_2019, mahnaz2025}, 
    \begin{align}
        \pauliGKP_{\paulivec} \coloneqq \op{D}\left( \paulivec \sqrt{\frac{\pi}{2}}\right) \projGKP = \projGKP \op{D}\left( \paulivec \sqrt{\frac{\pi}{2}}\right).
    \end{align} 
A useful representation for them is found using Eqs.~\eqref{eq:Paulishift} and~\eqref{eq:GKPprojector} and combining the displacements using $\op{D}(\vec{x})\op{D}(\vec{x}') = e^{i \vec{x}^\tp \mat{\Omega} \vec{x}'} \op{D}(\vec{x}+\vec{x}')$ to get
    \begin{align} \label{eq:Pauli_GKP_Subspace}
        \pauliGKP_{\paulivec} 
        & = \frac{1}{\sqrt{\pi}} \sum_{\vec{n}\in\mathbb{Z}^2} \op{D}\left( (2\vec{n}+\paulivec)\sqrt{\frac{\pi}{2}} \right)
        e^{i\pi \vec{n}^\tp \mat{\Omega} \paulivec}.
    \end{align}
The phase is always $\pm 1$ and satisfies $e^{i\pi \vec{n}^\tp \mat{\Omega} \paulivec} = e^{i\pi \paulivec^\tp \mat{\Omega} \vec{n}}$, as it must. The subspace Paulis have support only in the GKP subspace, in contrast to the logical shift operators in Eq.~\eqref{eq:logical-shifts}. To highlight this distinction, we use an overline to indicate operators acting only in the GKP subspace. In contrast, a standard hat indicates operators on the full CV Hilbert space, \emph{e.g.,} $\bar{A}$ vs $\hat{A}$.

GKP states require a finite energy constraint to be physical and normalizable, and this constraint is not unique~\cite{matsuura_equivalence_2020}. In this work, we focus on approximate GKP states described by applying the damping operator \cite{menicucci_fault-tolerant_2014}
        $e^{-\beta\op{n}}$
to ideal states, 
    $\ket{\Tilde{\psi}} \coloneqq \frac{1}{\sqrt{\mathcal{N}}} e^{-\beta\op{n}} \ket{\bar{\psi}}$,
where $\mathcal{N}$ is the normalization. 
Damping replaces the infinite Dirac-$\delta$ spikes in the ideal GKP position (and momentum) wavefunction with narrow Gaussians while simultaneously applying a broad Gaussian to damp spikes far from the origin\footnote{Note this is different from applying a broad envelope and replacing the spikes one after another, see \cite{matsuura_equivalence_2020}.}~\cite{menicucci_fault-tolerant_2014, mensen_phase-space_2021}. 
This spoils the orthogonality of approximate GKP states, leading to inevitable logical errors.
For small damping, $\beta$ is related to the GKP effective squeezing $\Delta_\text{dB}$~\cite{terhal2017sensor}, expressed in decibels (dB), via
    $\Delta_\text{dB} \approx -10\log_{10}\beta.$ 
An important value we make use of below is $\beta = 0.1$, which corresponds to a 10dB state, roughly the quality required for fault-tolerant operation~\cite{bourassa_blueprint_2021, tzitrin2021passive, walshe2025}.

\subsection{GKP error correction}\label{sec:2.9GKP_error_correction}

GKP error correction converts CV noise into qubit errors to be corrected~\cite{gottesman_encoding_2001,glancy_error_2006}. 
Nondestructively measuring the modular quadratures $\op{q} \text{ mod } \sqrt{\pi}$ and $\op{p} \text{ mod } \sqrt{\pi}$ projects the input state into an error subspace, which is then restored to the GKP subspace by applying corrective shifts. One way to do this involves coupling the noisy state to ancillae prepared in ideal GKP states that are measured via homodyne detection, a procedure called Steane error correction \cite{steane_active_1997}. 
Each measurement yields an outcome $m \in \mathbb{R}$, which we split into a (centered) integer multiple of $\sqrt{\pi}$ and a remainder~\cite{pantaleoni2021subsystem}
\begin{align}
    m = \lfloor m \rfloor_{\sqrt{\pi}} + \{m\}_{\sqrt{\pi}},
\end{align}
using the definitions
\begin{align}
    \lfloor m \rfloor_{\sqrt{\pi}} &\coloneqq \sqrt{\pi} \left \lfloor\frac{m}{ \sqrt{\pi}} + \frac{1}{2} \right\rfloor, 
    \\
    \{m\}_{\sqrt{\pi}} &\coloneqq m - \lfloor m \rfloor_{\sqrt{\pi}},
\end{align}
with $ \lfloor m \rfloor_{\sqrt{\pi}} \in \sqrt{\pi}\mathbb{Z}$ and $\{m\}_{\sqrt{\pi}} \in \left[-\frac{\sqrt{\pi}}{2},\frac{\sqrt{\pi}}{2}\right)$. Henceforth, we drop the subscripts for brevity; \emph{e.g.,} $\lfloor \cdot \rfloor_{\sqrt{\pi}} \rightarrow \lfloor \cdot \rfloor,$ being careful to remember this is not a simple floor function. 
Due to the periodicity of the ancillae, this measurement projects the state into an eigenstate of the modular quadrature with eigenvalue $\{m\}$. To return to the GKP subspace, a correction is applied to shift back to the $\{m\} = 0$ subspace of both quadratures. There are many choices of shift to accomplish this; shifting to the nearest integer is known as \emph{standard binning}.

A circuit to perform modular position measurement with a standard-binning correction employs a CV controlled X gate, $\op{C}_X^{12}=e^{-i\op{q}_1\op{p}_2}$, to provide the entanglement with GKP ancilla $\ket{\bar{+}}$,\footnote{
Circuits run right-to-left just as in Refs.~\cite{walshe_continuous-variable_2020, walshe_streamlined_2021,walshe_equivalent_2023}. This choice enables us to translate easily between the circuit and Kraus operators. \par} 
    \begin{equation} 
    \begin{split} \label{Cir:error_correction_pos}
    \Qcircuit @C=1.5em @R=1.4em
    {
    \lstick{(out)}&\gate{\op{X}^\dagger(\{m_q\})}\cwx[1] &\ctrl{1} &\rstick{(in)}  \qw \\
    &\nogate{\brasub{m_q}{q}} &\targ{} &\rstick{\ket{\overline{+}}} \qw
    } 
    \end{split}
    \qquad
    \end{equation}
After measurement, the input is corrected with a position shift,
    \begin{align}\label{eq:x_realshift}
        \op{X}(s) \coloneqq \op{D}\left(\tfrac{1}{\sqrt{2}}(s,0)^\tp \right) = e^{-i s \op{p}},
    \end{align}
by the modular part of the outcome, $s=-\{ m_q \}$. Note that this shift differs from $\op{X}_{L}$, Eq.~\eqref{eq:Paulishift}, which shifts in position by $\sqrt{\pi}$. 
The circuit for modular momentum measurement is constructed similarly, with the corrective momentum shift, 
\begin{align}\label{eq:z_realshift}
\op{Z}(t) \coloneqq \op{D}\left(\tfrac{1}{\sqrt{2}}(0,t)^\tp \right) = e^{i t \op{q}}, 
\end{align}
determined by the modular part of that outcome $t=-\{m_p\}$.
As above, $\op{Z}(-\{m_p\})$ differs from $\op{Z}_L$, Eq.~\eqref{eq:Paulishift}, which shifts by $\sqrt{\pi}$. Taken together, the two circuits perform GKP error correction [see Fig.~\ref{fig:circuit_table}(b) below].

The Kraus operator for the Steane GKP error-correction circuit is~\cite{baragiola_all-gaussian_2019, mensen_phase-space_2021, mahnaz2025}
\begin{align}
    \krausEC(\vec{m}) 
    &= 
    \op{Z}(\lfloor m_p \rfloor)\op{X}(\lfloor m_q \rfloor) \projGKP \op{Z}^\dagger(m_p)\op{X}^\dagger(m_q)
    \\
    &= \projGKP \op{Z}^\dagger( \{m_p\} )\op{X}^\dagger( \{m_q\} ) \label{eq:Krausop_GKP_correction}.
\end{align}
In the second line, we have pushed the $\sqrt{\pi}$-integer shifts through the GKP projector (they commute) and combined them with the shifts on the right-hand side. We ignore the trivial phase $e^{-i\{m_p\}\lfloor m_q \rfloor}$, as these Kraus operators never appear in linear combinations.

We define the GKP error correction channel as the average over syndrome outcomes,
    \begin{align} 
        \mathcal{E}_\text{EC} \coloneqq \int_{-\infty}^{\infty} d^2m\, \op{K}_\text{EC}(\vec{m})\odot\op{K}_\text{EC}^\dagger(\vec{m}).
\end{align}
Just as we split a real number into an integer and the remainder, we rewrite the integral as a sum over the integers and an integral over the remainder,
    \begin{align}
        \int_{-\infty}^{\infty}d^2m 
        =
        \sum_{\lfloor \vec{m} \rfloor \in \mathbb{Z}^2} \int_{-\sqrt{\pi}/2}^{\sqrt{\pi}/2}d^2 \{m\}.
\end{align}
As a consequence of standard binning, the Kraus operator in Eq.~\eqref{eq:Krausop_GKP_correction} does not depend on the integer part of the syndromes, $\lfloor \vec{m} \rfloor$, and therefore, we need to integrate over only the remainders~\cite{mahnaz2025},
    \begin{align}\label{eq:GKPerrorcor}
        \mathcal{E}_\text{EC} \propto \int_{-\sqrt{\pi}/2}^{\sqrt{\pi}/2} d^2\{m\} \, \krausEC(\vec{m}) \odot \krausEC^\dagger(\vec{m}).
    \end{align}

An equivalent method to perform GKP error correction, known as Knill error correction \cite{knill_quantum_2005}, involves teleportation of the noisy state through a GKP Bell pair. Knill error correction with controlled gates is equivalent to Steane error correction---in the ideal regime---with controlled gates~\cite{mahnaz2025}, and it can also be implemented with beam splitters~\cite{walshe_continuous-variable_2020}, making it useful in optics. We focus on Steane error correction here; these equivalences mean our results directly carry over to Knill error correction.

\begin{figure}[b]
    \includegraphics[width=0.35\textwidth]{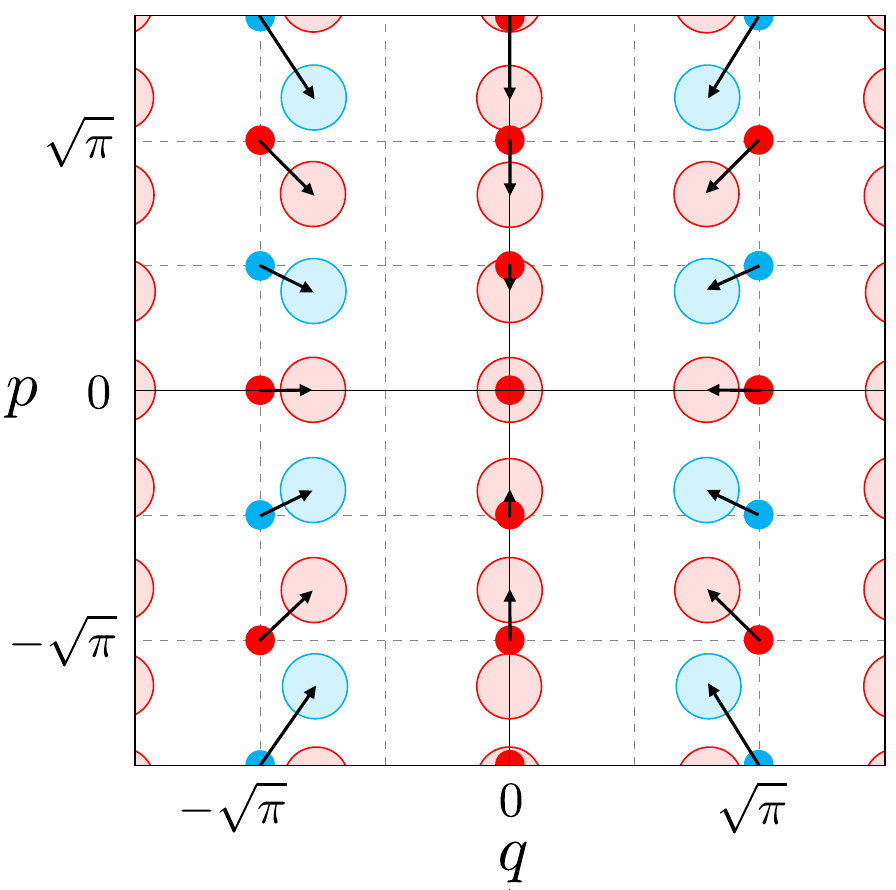}
    \caption[]{\label{fig:losscartoon}Effects of loss on a GKP state. Solid dots indicate spikes in the Wigner function of a high-quality GKP $\ket{\bar{0}}$ state. Larger circles show the evolution of each spike under pure loss---it is drawn towards the origin with added noise.    
    }
\end{figure}

\section{Pure loss}\label{sec:2.9Pure-loss}

The dominant source of noise in many systems is \emph{pure loss} \cite{knill_scheme_2001}, which generically describes the loss of excitations. In optical systems, pure loss describes photons scattered out of the mode of interest. Scattering from waveguides or optical fibres, mode mismatch when coupling on and off chips and fibres, and unwanted reflections from beam splitters~\cite{eisert_gaussian_2005} are significant contributors. Pure loss at rate $0 \leq\gamma \leq 1$ causes a contraction in phase space $(\op{q}, \op{p} ) \rightarrow ( \sqrt{1-\gamma}\op{q}, \sqrt{1-\gamma}\op{p} )$ that does not act uniformly; points farther from the origin experience a more significant shift towards the origin, let us clarify.

 When loss acts on a damped GKP state, the peaks experience two effects: (1) They are drawn towards the origin, and (2) their individual peak variances are increased from $\frac{1}{2}\tanh{\beta}$ to $\frac{1}{2}\tanh{\beta}(1-\gamma)+\frac{\gamma}{2}$ \cite{hastrup_analysis_2023}. Together, these effects contribute to logical errors in decoding by pushing the probability into quadrature measurement bins associated with logical errors. 
Furthermore, these effects disproportionately impact higher-quality GKP states, whose distant peaks face greater drag, thereby increasing error rates. An illustration of this effect is shown in Fig.~\ref{fig:losscartoon}.

A standard protocol to deal with the effects of loss is to either pre-amplify or post-amplify the state with phase-insensitive amplification~\cite{noh_quantum_2019,fukui2021alloptical}. When this amplification is appropriately tuned to the level of loss, the resulting noise is a Gaussian random displacement channel. This protocol has been useful in analyzing loss; however, the amplification inevitably introduces more noise. In fact, in the regime of high quality but not unrealistic damping (under 15 dB), it has been shown that GKP error correction performs best without amplification~\cite{hastrup_analysis_2023,shaw_stabilizer_2024}. Further work has shown that an optimal recovery procedure using the transpose channel 
also does not require amplification~\cite{zheng_performance_2024}. In this work, we focus on ideal GKP error correction with standard binning to derive qubit channels. Future work could expand this study to include other decoding strategies. 
Next, we provide two useful representations for the pure-loss channel based on two physical unravellings that we will employ when deriving a logical GKP qubit channel below.

\subsection{Kraus representations of the loss channel} \label{sec:representations_losschannel}

Regardless of its origin in an optical system, pure loss at a rate of $0 \leq \gamma \leq 1$ can be modeled via the Stinespring dilation, 
\begin{equation} 
\begin{split} \label{Cir:loss_circuit}
    \Qcircuit @C=2.0em @R=2.5em  
    {
    &&&\lstick{(out)}  & \varbs{1} & \rstick{(in)}  \qw \\
    &&&\lstick{//}  & \qw & \rstick{\ket{0}} \qw
    } 
\end{split}
\qquad\qquad,
\end{equation}
where $``//"$ indicates the trace, and the beam splitter is defined as \cite{walshe_continuous-variable_2020}
\begin{equation}
\begin{split} \label{cir:beamsplitter}
	\raisebox{-1.2em}{$\op{B}_{jk}(\theta)  \coloneqq
 e^{ - i\theta ( \op{q}_j \op{p}_k - \op{p}_j \op{q}_k )} \, = \, $~~}
         \Qcircuit @C=1.25em @R=2.5em @! 
         {
         	& \varbs{1} & \rstick{j}  \qw \\
         	& \qw       & \rstick{k} \qw
  		  } 
\end{split} \quad\quad. 	
\end{equation}
The loss rate is given by the beam-splitter reflectivity $\gamma = R = s_\theta^2$, where we 
introduce abbreviated notation for the trigonometric functions,
    \begin{subequations} \label{eq:trig_to_gamma}
    \begin{align}
        s_\theta & \coloneqq \sin \theta = \sqrt{\gamma},
        \\
        c_\theta &\coloneqq \cos \theta = \sqrt{1-\gamma},
        \\
        t_\theta &\coloneqq \tan \theta = \sqrt{\frac{\gamma}{1-\gamma}} \, .
    \end{align}
    \end{subequations}

We consider the trace in circuit~\eqref{Cir:loss_circuit} to be a measurement of the ancilla mode in some basis followed by a discarding of the outcome. Each outcome $j$ corresponds to a Kraus operator $\op{L}_j$ in a representation of the channel
    \begin{align} \label{eq:newnoise}
        \mathcal{E}_\text{loss} = \sumint_j \op{L}_j \odot \op{L}_j^\dagger ,
    \end{align}    
where $\sumint$ indicates that our Kraus operators can be discrete or continuous depending on the measurement basis. For completely-positive trace-preserving (CPTP) channels, such as loss, the Kraus operators resolve the identity $\sumint_j \op{L}_j^\dagger \op{L}_j = \op{I}_\text{CV}$, and the probability (or probability density) of outcome $j$ is given by $\Pr(j) = \Tr[\op{L}_j^\dagger \op{L}_j \op{\rho}]$, where $\op{\rho}$ is a state in the input mode.
Below, we consider two measurements: photon counting, which gives a standard discrete representation for pure loss, and heterodyne detection, which offers a new representation that proves useful in the calculations that follow. In Appendix~\ref{Appendix:losschannelhomodyne}, we give another novel description of the pure-loss channel arising from homodyne measurement.

\subsubsection{Discrete representation from photon counting}

Consider photon counting of the ancillary mode with outcome $j \in \mathbb{Z}_{\geq 0}$, corresponding to projection onto Fock state $\ket{j}$. The circuit for this situation is
\begin{equation} 
\begin{split} \label{Cir:loss_circuit_d}
    \Qcircuit @C=2.0em @R=2.5em  
    {
    &&&\lstick{(out)}  & \varbs{1} & \rstick{(in)}  \qw \\
    &&&\lstick{\bra{j}}  & \qw & \rstick{\ket{0}} \qw
    } 
\end{split}
\qquad\qquad ,
\end{equation}
where, due to the right-to-left convention of our circuits, they have the same operation ordering as their resulting Kraus operators, removing the need to rearrange operators like in the traditional left-to-right circuits. 
The Kraus operator for this circuit is
    \begin{align} \label{eq:loss_KrausFock}
        \krauslossphoton_j
            = t^j_\theta \frac{\op{a}^j}{\sqrt{j!}} c^{\op{n}}_\theta 
            = \left( \frac{\gamma}{1-\gamma} \right)^{\frac{j}{2}} \frac{\op{a}^j}{\sqrt{j!}} (1-\gamma)^{\frac{\op{n}}{2}} ,
    \end{align}
with the superscript P indicating photon counting; the derivation of this Kraus operator is contained in Appendix.~\ref{Appendix:losschannelpho}.
Fock states resolve the identity $\op{I}_\text{CV} =\sum_{j=0}^\infty \outprod{j}{j}$, giving a discrete representation of the loss channel,
    \begin{align} \label{eq:loss-discrete-rep}
        \mathcal{E}_\text{loss} \coloneqq \sum_{j=0}^\infty \krauslossphoton_j \odot \big( \krauslossphoton_j \big)^\dagger.
    \end{align}

Since the ancilla mode is initially in vacuum, one might imagine that outcome $j$ counts the number of photons lost from the input state. However, this is true only when the reflectivity $\gamma$ is close to zero 
\cite{weedbrook_gaussian_2012, albert_performance_2018, noh_quantum_2019, hastrup_analysis_2023}. This caveat is due to the presence of the damping operator $(1-\gamma)^{\frac{\op{n}}{2}}$ in the loss-channel Kraus operators. For large $\gamma$, the damping suppresses support on high photon number in the input state. This reduces the probability for 
high photon counts (large $j$)
even before the subtraction $\op{a}^j$ is applied. 

When no photons are counted, $j=0$, the Kraus operator 
    \begin{align} \label{eq:envelope_from_kraus}
        \krauslossphoton_{j=0} &= c^{\op{n}}_\theta = (1-\gamma)^{\frac{\op{n}}{2}}
    \end{align}
is equivalent to a damping operator, $e^{-\beta\op{n}}$, with $\beta = -\ln{c_\theta} = \ln{ (1 -\gamma)^{-\frac{1}{2}}}$. This effective damping appears in every Kraus operator in Eq.~\eqref{eq:loss_KrausFock}. A
non-zero $j$ corresponds to a process referred to as ``photon subtraction'' that is relevant when $\theta \ll 1$, and forms the foundation for various quantum-optical protocols, notably when a single photon is detected. 
This is particularly interesting for GKP states, which lie in the even parity subspace~\cite{grimsmo_quantum_2020}. Subtraction of a single photon (or any odd number) projects a GKP state into the odd-parity subspace, making it completely orthogonal to the GKP subspace. We elaborate on photon subtraction further in Sec.~\ref{sec:photon_subtraction}.

\subsubsection{Continuous representation from heterodyne detection}

Consider heterodyne detection of the ancillary mode with outcome $\mu \in \mathbb{C}$, corresponding to projection onto coherent state $\ket{\mu}$. The circuit for this situation is
\begin{equation} 
\begin{split} \label{Cir:loss_circuit_c}
    \Qcircuit @C=2.0em @R=2.5em  
    {
    &&&\lstick{(out)}     &\varbs{1} &\rstick{(in)}  \qw \\
    &&&\lstick{\bra{\mu}} &\qw       &\rstick{\ket{0}} \qw
    } 
\end{split}
\qquad\qquad ,
\end{equation}
corresponding to the Kraus operator
    \begin{align} \label{eq:loss_KrausCoherent}
        \krauslosshet(\mu) 
        &= e^{-\frac{1}{2}|\mu|^2} e^{\mu^* t_{\theta} \op{a} } c_{\theta}^{\op{n}} \\
        &=  e^{-\frac{1}{2}|\mu|^2} e^{\mu^* \sqrt{\frac{\gamma}{1-\gamma}} \op{a} }  (1-\gamma)^{\frac{\op{n}}{2}}, 
    \end{align}
 with the superscript Het indicating heterodyne measurement. The derivation for this Kraus operator is contained in Appendix.~\ref{Appendix:losschannelhet}.
Coherent states resolve the identity as $\op{I}_\text{CV} = \frac{1}{\pi} \int d^2\mu \, \outprod{\mu}{\mu}$, giving a continuous representation of the loss channel
    \begin{align} \label{eq:loss-continuous-rep}
        \mathcal{E}_\text{loss} = \int_{-\infty}^{\infty} \frac{d^2\mu}{\pi}\krauslosshet(\mu) \odot \big[ \krauslosshet(\mu) \big]^\dagger.
    \end{align}

Just as in the case of the photon-counting measurement, if we set our measurement outcome to zero ($\mu=0$ in this case), Eq.~\eqref{eq:loss-continuous-rep} is identical to Eq.~\eqref{eq:envelope_from_kraus} and equivalent to damping. 

\subsubsection{Connecting the representations}

Using the Fock representation of coherent states,
  $  \ket{\alpha}=e^{-\frac{|\alpha|^2}{2}}\sum_{n=0}^\infty\frac{\alpha^n}{\sqrt{n!}}\ket{n},$
the Kraus operators in Eq.~\eqref{eq:loss_KrausCoherent} can be related to those derived from photon counting, Eq.~\eqref{eq:loss_KrausFock}, via
    \begin{align} \label{eq:lossc_to_lossf}
        \krauslosshet(\mu) = e^{-\frac{1}{2}|\mu|^2} \sum_{j=0}^\infty \frac{(\mu^*)^j}{\sqrt{j!}}\krauslossphoton_j.
\end{align}
The relationship between the Kraus operators in the other direction comes from 
    \begin{align}
        \ket{n} = \frac{1}{\sqrt{n!}} \partial^n_{\alpha} e^{\frac{|\alpha|^2}{2}} \ket{\alpha} \bigg|_{\alpha=0},
    \end{align}
where $\partial^n_\alpha$ is the $n$th partial derivative with respect to $\alpha$.
This gives the relation
\begin{align} \label{eq:lossf_to_lossc}
    \krauslossphoton_j = \frac{1}{\sqrt{j!}} \partial^j_{\mu^*} \left( e^{\frac{1}{2}|\mu|^2} \krauslosshet(\mu)\right) \bigg|_{\mu^*=0}.
\end{align}
Equations~\eqref{eq:lossc_to_lossf} and \eqref{eq:lossf_to_lossc} allow us to transform back and forth between representations of the loss channel, which will prove valuable in the derivations below.\footnote{The connections between the Kraus operators depend on the relationship between the number and heterodyne measurement bases but not on the structure of the ancilla state or the entangling operation.} 

\begin{figure*}
    \centering
    \begin{subfigure}
        \centering
        \includegraphics[width=\textwidth]{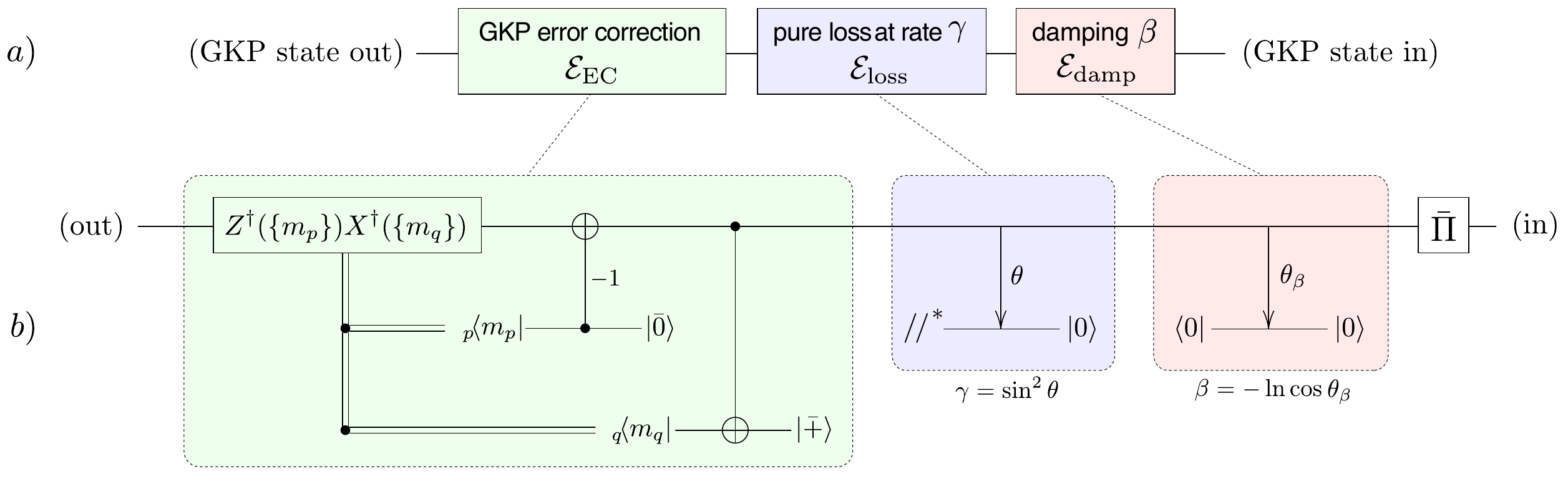}
    \end{subfigure}
    \begin{subfigure}
        \centering
        \renewcommand{\arraystretch}{1.5}  %
    \begin{tabular}{|c|c|c|c||c||}  %
        \hline
                          & No loss (damping only) & Heterodyne-heralded loss & Photon-heralded loss & Pure loss  \\ \hline
        Post-selected GKP syndrome & \makecell{Analytic, Eq.~\eqref{Blochvectorcomponents_dampingonly}  \\    Fig.~\ref{fig:damping_noloss}  } & Analytic, Eq.~\eqref{eq:heterodyne-heralded_Blochvector} & Analytic, Eq.~\eqref{eq:photoncounting-heralded-Blochvector}       & \makecell{ Analytic, Eq.~\eqref{eq:matrixelement_avgloss} \\ Fig.~\ref{Fig:processmatrix_loss} }          \\ \hline
        Average GKP syndrome & \makecell{Numeric \\ Fig.~\ref{fig:damping_noloss} }   & Numeric$^\dagger$       & \makecell{ Numeric \\ Fig.~\ref{fig:Blochsphere_heralded_loss} }       & \makecell{ Numeric \\ Fig.~\ref{Fig:processmatrix_loss} }                 \\ \hline
    \end{tabular}
    \end{subfigure}
    \hfill
    \caption{(a) We consider the following logical channel that maps ideal GKP states to ideal GKP states: An ideal GKP state undergoes damping of strength $0 \leq \beta \leq \infty$ followed by pure loss at rate $0 \leq \gamma \leq 1$. Then, ideal GKP error correction projects the damaged state back into the GKP subspace. (b) A realization of the composed channels, with damping and loss each described by a dilation using an ancilla mode prepared in vacuum coupled to the data mode with a beam splitter. The circuit runs right-to-left. Tracing over the ancilla mode for loss gives the pure loss channel. Further, we consider both single-syndrome error correction and error correction averaged over syndromes. We produce the $4 \times 4$ single-qubit process matrix describing the logical GKP qubit map for all cases. 
    (c) Table summarizing our results. We give either the Bloch vector components of the Kraus operator (from which the process matrix follows) or the process matrix elements themselves. Figures include depictions of the process matrices and Bloch vector trajectories given ideal GKP Pauli eigenstates.\\[0.5ex]
    ${}^*$~We treat the trace as projective measurements of the loss mode in two relevant bases corresponding to heterodyne and number detection, respectively, and then discarding the outcome. By retaining the outcome, this additionally allows us to find conditional qubit maps that depend on the measurement basis and are heralded by the outcome.\\[0.5ex]
    $^\dagger$~We focus on heralded photon-counting outcomes, since number-resolving detection is a key non-Gaussian operation employed in photon subtraction and Gaussian boson sampling.
    \label{fig:circuit_table}}
\end{figure*}

\section{GKP qubit channel for loss}\label{sec:2.9Results}
The combined effects of damping,
\begin{align}
\mathcal{E}_\text{damp} \coloneqq e^{-\beta\op{n}}\odot e^{-\beta\op{n}},
\end{align} 
pure loss $\mathcal{E}_\text{loss}$ [Eq.~\eqref{eq:newnoise}], and ideal GKP error correction $ \mathcal{E}_\text{EC}$ [Eq.~\eqref{eq:GKPerrorcor}] amount to a qubit channel $\qubitChannel$,
    \begin{align} \label{eq:qubitchannelcomposed}
        \qubitChannel = \mathcal{E}_\text{EC} \circ \mathcal{E}_\text{loss} \circ \mathcal{E}_\text{damp} \circ \projGKP,
    \end{align}
where, for completeness, we include a GKP projector at the start to ensure that this channel acts only in the two-dimensional subspace of the GKP code. 
The circuit depicting the composed channel is shown in Fig.~\ref{fig:circuit_table}.
In what follows, we are concerned with the effects of this channel on input ideal GKP states. The damping map $\mathcal{E}_\text{damp}$ is not trace-preserving on the CV Hilbert space---it is trace decreasing---and, as such, will not be trace-preserving in the qubit subspace either. Nevertheless, we use ``qubit channel'' in this work to refer to a completely positive map on the GKP subspace, with the understanding that ``channel'' is used in other places to refer only to CPTP maps~\cite{mahnaz2025}.

\begin{figure*}[hbt!]
    \includegraphics[width=1.0\textwidth]{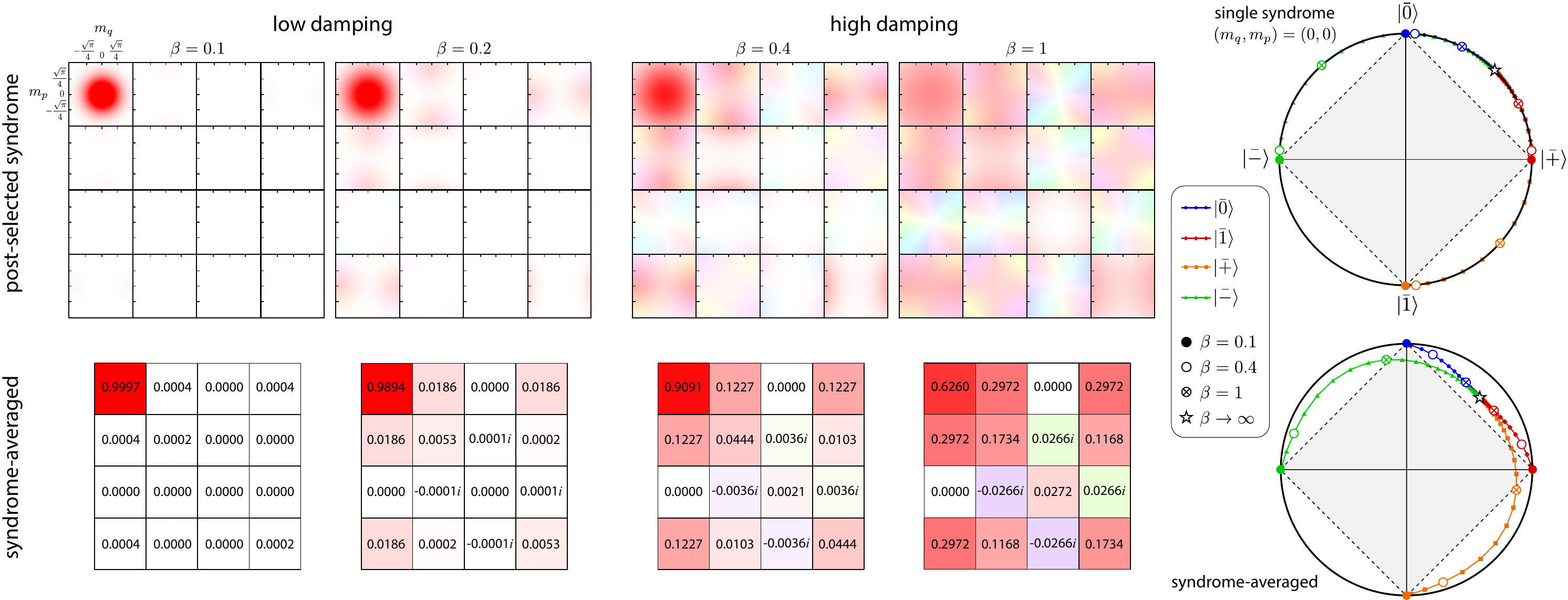}
    \caption[]{\label{fig:damping_noloss}
    Process matrices and Bloch vectors for no loss ($\gamma = 0$) and various damping parameters.
    Top row: each subplot is a plot of the process matrix element $\chi_{\paulivec \paulivec'}(\vec{m})$ conditioned on syndrome outcomes $\{m_q\}$ and $\{m_p\}$. See Fig.~\ref{fig:legend} for a guide to reading these plots and a color legend.
    Bottom row: syndrome-averaged process matrices, normalized to their trace, for each value of $\beta$. To the right are trajectories in the XZ plane of the qubit Bloch sphere as a function of $\beta$ for GKP Pauli $X/Z$ eigenstates with damping. The top right plot shows Bloch vectors conditional on syndrome $(0,0)$. The bottom right plot shows Bloch vectors averaged over syndromes. This plot reproduces curves found in Shaw \emph{et al.}~\cite{shaw_stabilizer_2024}. Highlighted on each are low damping (solid circles), medium damping (open circles), and high damping (crossed circles). In each plot, a star indicates the fixed point of maximum damping. For the syndrome $(0,0)$, this state is $\ket{+H}$; for the syndrome-averaged case, this state is given by Eq.~\eqref{eq:avgECvacstate}.
    }
\end{figure*}

\begin{figure}[b!]
~\\
    \includegraphics[width=0.5\textwidth]{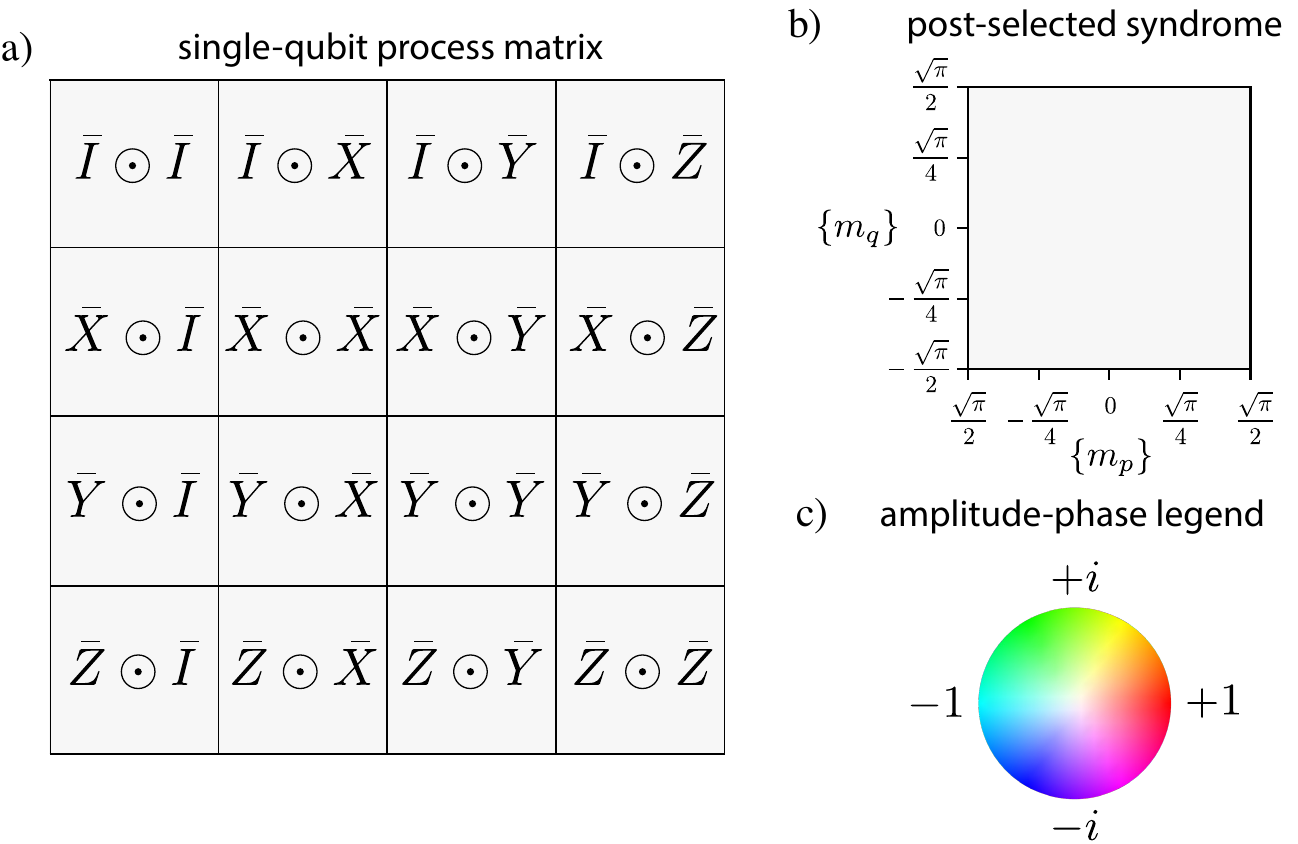}
    \caption[]{\label{fig:legend}Guide to read the process-matrix figures. (a) We represent the matrix elements of a single-qubit process matrix $\mat{\chi}$ as subplots. The plots are laid out in an array just as if they were elements of a matrix. 
    For syndrome-averaged process matrices, the values in each subplot are summed and then normalized by the trace of the syndrome-averaged process matrix, $\chi_{\paulivec \paulivec'} \rightarrow \chi_{\paulivec \paulivec'}/\sum_{\paulivec} \chi_{\paulivec \paulivec}$, and then plotted. The matrix-element values are overlaid as a final step.
    (b) For post-selected syndromes, each subplot is a plot of $\chi_{\paulivec\paulivec'}(\vec{m})$ over the region of fractional syndromes $[-\frac{\sqrt{\pi}}{2}, \frac{\sqrt{\pi}}{2} )\times [-\frac{\sqrt{\pi}}{2}, \frac{\sqrt{\pi}}{2} )$. For visual comparison, these plots are also normalized to the trace of the syndrome-averaged process matrix.
    (c) In all plots, phase is given by color, and amplitude is given by brightness according to the color wheel.
    }
\end{figure}

Kraus representations for the qubit channel, 
    \begin{align}
        \qubitChannel = \sumint_{\vec \ell} \qubitKraus(\vec{\ell}) \odot \big[ \qubitKraus(\vec{\ell}) \big]^\dagger ,
    \end{align} 
are immediately available by combining the Kraus operators of each composed channel in Eq.~\eqref{eq:qubitchannelcomposed}, where $\vec{\ell}$ is a placeholder for the relevant variables: error correction outcomes and loss-heralding outcome. A single qubit channel (or map) is also described in a basis of Paulis as
    \begin{align}\label{eq:qubitChan_def}
         \qubitChannel = 
        \sum_{\paulivec, \paulivec'}  \chi_{\paulivec \paulivec'} \, \pauliGKP_{\paulivec} \odot \pauliGKP_{\paulivec'} \, ,        
    \end{align}
where the Hermitian, positive-semi-definite matrix $\mat{\chi} \in \mathbb{R}^{4 \times 4}$ is called the \emph{process matrix}.
Projecting the Kraus operators onto the GKP subspace Paulis, Eq.~\eqref{eq:Pauli_GKP_Subspace}, $\qubitKraus(\vec{\ell}) = \frac{1}{2} \sum_{\paulivec} r_{\paulivec}(\vec{\ell}) \, \pauliGKP_{\paulivec}$,
gives four GKP Bloch-vector components,\footnote{ The factors of $\frac{1}{2}$ in the Kraus operator and $\frac{1}{4}$ in the $\chi$ matrix come from the fact that the Paulis are not normalized; $\Tr [\pauliGKP_{\paulivec}^2] = 2$ up to an infinite constant.} 
    \begin{align} \label{eq:Blochvectorcomponent}
        r_{\paulivec}(\vec{\ell}) \coloneqq \Tr[\qubitKraus(\vec{\ell}) \pauliGKP_{\paulivec} ].
    \end{align}
The process-matrix representation of the qubit map conditional on $\vec{\ell}$ follows directly, 
    \begin{align}
        \qubitKraus(\vec{\ell}) \odot \left[\qubitKraus(\vec{\ell})\right]^\dagger = 
        \sum_{\paulivec, \paulivec'} \chi_{\paulivec \paulivec'}(\vec{\ell}) \, \pauliGKP_{\paulivec} \odot \pauliGKP_{\paulivec'} \, ,
    \end{align}
with matrix elements 
    \begin{align} \label{eq:process-matrix-elements}
        \chi_{\paulivec \paulivec'}(\vec{\ell}) =  \frac{1}{4}  r_{\paulivec}(\vec{\ell}) r^*_{\paulivec'}(\vec{\ell}).
    \end{align}
Summing over all relevant variables gives a $\chi$-matrix representation for the GKP qubit channel in Eq.~\eqref{eq:qubitchannelcomposed} with process-matrix elements
    \begin{align}
        \chi_{\paulivec\paulivec'} = \sumint_{\vec{\ell}} \chi_{\paulivec\paulivec'}(\vec{\ell}).
    \end{align}
One is also free to sum over only some of the variables in $\vec{\ell}$, giving process matrices for other important physical settings. 
    
We summarize the settings of interest and our results in Fig.~\ref{fig:circuit_table}. Post selecting on a single GKP syndrome $\vec{m}$, we find analytic forms for the Kraus operators describing the qubit map given (a) no loss, (b) heterodyne detection of the loss mode, (c) homodyne detection of the loss mode, and (d) the pure loss channel. Taking the average over error correction syndromes is also possible, but the resulting integrals of $\Theta$ functions do not lend themselves to useful analytic forms. Thus, syndrome averaging for each case is performed numerically as described in the caption of Fig.~\ref{fig:legend}. Below, we discuss each physical setting in detail and present our analytic results.

\subsection{Heralded loss} \label{sec:heraldedloss-postselectedEC}

    We begin by considering the conditional evolution generated by a measurement of the loss mode followed by GKP error correction. These dynamics are conditional on both the loss-mode outcome---$\mu$ for heterodyne measurement or $j$ for photon counting---and the GKP error correction outcomes $\vec{m}$. Later, we can average over one or both of them to analyze average evolutions.

    \subsubsection{Heterodyne measurement of the loss mode}

Using the continuous representation of loss, Eq.~\eqref{eq:loss-continuous-rep}, we can write Eq.~\eqref{eq:qubitchannelcomposed} as 
    \begin{align}   
        \qubitChannel = \int_{-\sqrt{\pi}/2}^{\sqrt{\pi}/2} d^2\{m\} \int \frac{d^2 \mu}{\pi} \qubitKraushet(\mu,\vec{m}) \odot \big[ \qubitKraushet(\mu,\vec{m}) \big]^\dagger,
    \end{align}
with qubit Kraus operators
    \begin{align} \label{eq:hetecKraus}
        \qubitKraushet(\mu,\vec{m}) = \krausEC(\vec{m}) \krauslosshet (\mu) \op{N}_{\beta} \projGKP .
    \end{align}
The GKP Bloch vector components, Eq.~\eqref{eq:Blochvectorcomponent}, are found by inserting the definitions of the operators above,
using the displacement-operator representation of $\pauliGKP_{\paulivec}$, Eq.~\eqref{eq:Pauli_GKP_Subspace}, and taking the trace in the coherent-state basis. After a lengthy calculation, presented in Appendix~\ref{Appendix:ProcessMatrixDerivation}, we find the analytic form
    \begin{align} \label{eq:heterodyne-heralded_Blochvector}
        r^\het_{\paulivec}(\mu,\vec{m}) 
        &\circeq\frac{ e^{\beta} e^{-\frac{1}{2}|\mu|^2} e^{\pi i (\vec{w}_{\paulivec}^{-})^\tp \mat{\Omega} \vec{w}_{\paulivec}^{+} } }{ \sqrt{\pi} ( c_\theta + e^\beta )}  
        \Theta \begin{bmatrix} - \mat{\Omega} \vec{w}_{\paulivec}^{+} \\ \vec{w}_{\paulivec}^{-} \end{bmatrix} \left(\vec{z}^\het; \mat{\tau}^\het \right) 
    \end{align}
in terms of a Siegel $\Theta$ function,
\begin{align}
    \Theta
    \begin{bmatrix}
    \vec{v}_1\\
    \vec{v}_2
    \end{bmatrix}(\vec{z};\mat{\tau}) 
    \coloneqq 
    \sum_{\vec{n}\in\mathbb{Z}^2} e^{2\pi i \left[\frac{1}{2}(\vec{n}+\vec{v}_1)^\tp \mat{\tau}(\vec{n}+\vec{v}_1)+(\vec{n}+\vec{v}_1)^\tp (\vec{z}+\vec{v}_2)\right]},
\end{align}
with parameters 
    \begin{subequations} \label{eq:params_dampherald}
    \begin{align}
        \mat{\tau}^\het &= i\tanh \left[ \tfrac{1}{2}(\beta-\ln{c_\theta}) \right] \mat{I} , \label{eq:taumat_dampherald}
        \\ 
        \vec{z}^\het &= -\frac{s_{\theta}\mu^*}{\sqrt{2\pi} (c_\theta + e^\beta)} \begin{bmatrix} 1 \\ i \end{bmatrix} ,
        \label{eq:zmat_dampherald}
        \\
        \vec{w}_{\paulivec}^{\pm} &= \frac{1}{2} \left( \paulivec\pm \frac{\{\vec{m}\}}{\sqrt{\pi}} \right),
        \label{eq:vec_z}
    \end{align}
    \end{subequations}
where $\mat{\Omega}$ is the two-dimensional symplectic form in Eq.~\eqref{eq:Pauli_GKP_Subspace}, $\mat{I}$ is the two-dimensional identity matrix, and the equality $\circeq$ is up to the global phase $e^{\frac{i}{2}\{m_q\}\{m_p\}}$.
The expression can also be written in terms of the loss parameter $\gamma$ using Eqs.~\eqref{eq:trig_to_gamma} and then manipulated in various ways using the properties of Siegel $\Theta$ functions (see Appendix~\ref{Appendix.2.9theta_function}). Importantly, the $\mat{\tau}$ matrix in Eq.~\eqref{eq:taumat_dampherald} is diagonal, allowing the Siegel $\Theta$ function to be split into two single-variable Jacobi $\theta$ functions, which are more swiftly evaluated than Siegel $\Theta$ functions in some software packages. 

From the expression above for the Bloch-vector components of $\qubitKraushet(\mu, \vec{m})$, the process matrix elements follow directly from Eq.~\eqref{eq:process-matrix-elements},
        $\chi^\het_{\paulivec,\paulivec'}(\mu,\vec{m}) = 
        \frac{1}{4}
        r^\het_{\paulivec}         
        (\mu,\vec{m}) \big[ r^{\het}_{\paulivec'}(\mu,\vec{m}) \big]^*.$
They, too, can be written as a single, four-dimensional Siegel $\Theta$ function or a product of four one-dimensional Jacobi $\Theta$ functions.

\subsubsection{Photon counting of the loss mode} \label{subsection:photoncounting}

Using the discrete representation of loss, Eq.~\eqref{eq:loss-discrete-rep}, we can write Eq.~\eqref{eq:qubitchannelcomposed} as 
    \begin{align}   
        \qubitChannel =  \int_{-\sqrt{\pi}/2}^{\sqrt{\pi}/2} d^2\{m\} \sum_j \qubitKrausphoton_j(\vec{m}) \odot \big[ \qubitKrausphoton_j(\vec{m}) \big]^\dagger,
    \end{align}
with qubit Kraus operators
    \begin{align}
        \qubitKrausphoton_j(\vec{m}) = \krausEC(\vec{m}) \krauslossphoton_j(\mu) \op{N}_{\beta} \projGKP .
    \end{align}
To find the Bloch-vector components, we transform directly from those for heterodyne detection, Eq.~\eqref{eq:heterodyne-heralded_Blochvector}, using the relation between the Kraus operators in Eq.~\eqref{eq:lossf_to_lossc}. Formally, this gives
\begin{align} \label{eq:photoncounting-heralded-Blochvector}
    r^\photon_{\paulivec}(j,\vec{m}) 
    &\circeq \frac{1}{\sqrt{j!}} \partial^j_{\mu^*} e^{\frac{1}{2}|\mu|^2} r^\het_{\paulivec}(\mu,\vec{m}) \bigg|_{\mu^*=0} ,
\end{align}
giving process matrix elements $\chi^\photon_{\paulivec,\paulivec'}(j,\vec{m}) =
 \frac{1}{4}  
r^\photon_{\paulivec}(j,\vec{m}) \big[ r^{\photon}_{\paulivec}(j,\vec{m}) \big]^*$. We found that this procedure is simpler than deriving the photon-counting Bloch vector directly. Although the derivatives above can be performed analytically in the infinite-sum representation of the Siegel $\Theta$ functions, the resulting terms cannot be resummed into known functions, as far as we can tell. In practice, we evaluate these terms using numerical derivatives of the Siegel $\Theta$ functions. 

\subsubsection{Special cases: No loss and complete loss}

In the case where loss is low, $\ln \sqrt{1-\gamma} \ll \beta $, the qubit dynamics are dominated by noise from the damping operator. In the limiting case of no loss, the logical channel is
  \begin{align} \label{eq:qubitchannel_noloss}
        \qubitChannel = \mathcal{E}_\text{EC} \circ \mathcal{E}_\text{damp} \circ \projGKP .
    \end{align}
The Bloch-vector components for this case can be found directly from those for heralded heterodyne detection, Eq.~\eqref{eq:heterodyne-heralded_Blochvector}. Taking the loss parameter to zero, $\gamma = 0$, gives $s_\theta = 0$ and $c_\theta = 1$, giving Siegel $\Theta$ function parameters
    \begin{subequations} \label{Blochvectorcomponents_dampingonly}
    \begin{align}
        \mat{\tau}^\het &\rightarrow i\tanh  \left(\tfrac{1}{2}\beta \right) \mat{I}, 
        \\
        \vec{z}^\het &\rightarrow \vec{0}.
    \end{align}
    \end{subequations}
Since damping is a coherent noise process, the qubit channel for large $\beta$ becomes exceedingly non-Pauli, even when averaged over syndromes. This behavior is apparent in  Fig.~\ref{fig:damping_noloss}, with non-Pauli features becoming more evident as $\beta$ increases. In the regime of fault tolerance, $\beta \leq 0.1$, damping acts primarily as the identity channel.
\begin{figure*}[t]
    \includegraphics[width=0.95\textwidth]{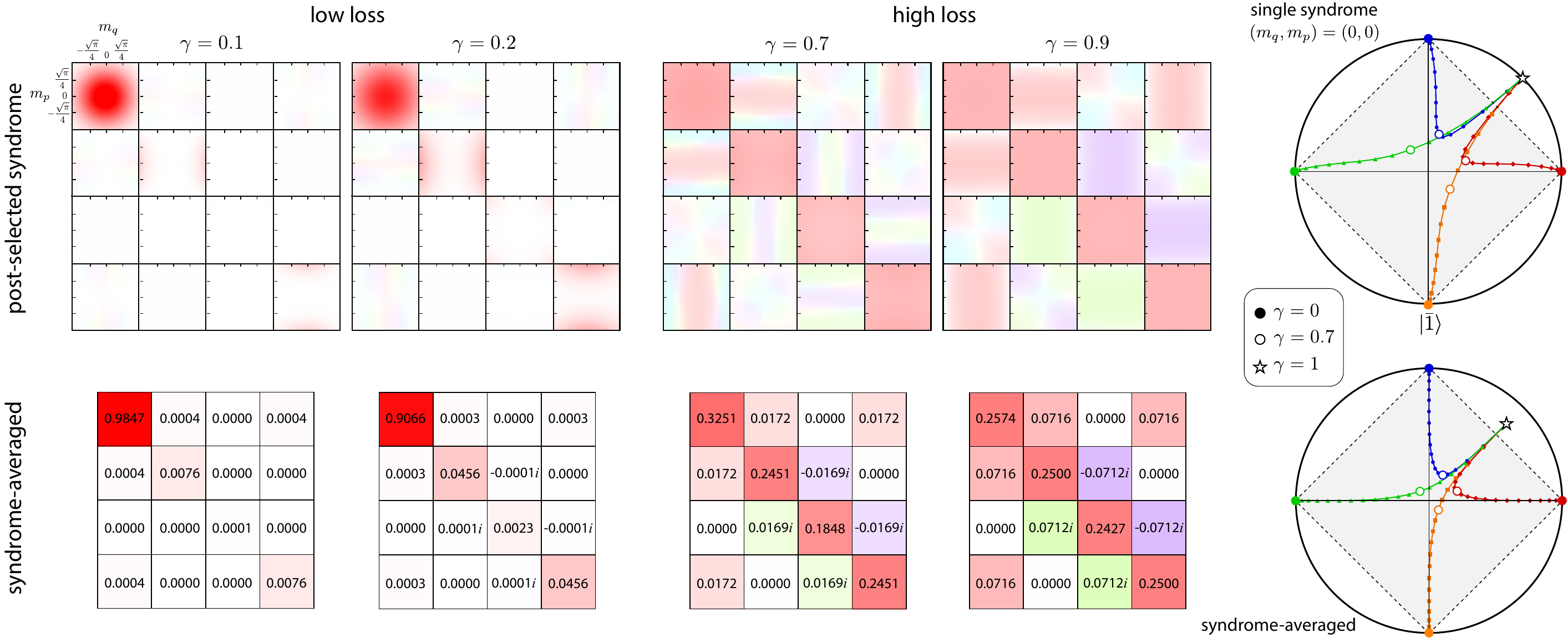}
    \caption[]{\label{Fig:processmatrix_loss} 
    Process matrices and Bloch vectors for fixed damping ($\beta = 0.1$) and various loss rates.
    See Fig.~\ref{fig:legend} for a guide to reading these plots and a color legend.
    Top row: Process matrices conditioned on syndrome outcomes $\{m_q\}$ and $\{m_p\}$. 
    Bottom row: syndrome-averaged process matrices for each value of loss parameter $\gamma$. To the right are trajectories in the XZ plane of the qubit Bloch sphere as a function of $\gamma$ for GKP Pauli $X/Z$ eigenstates. Top: Bloch vectors conditional on syndrome $(0,0)$. Bottom: syndrome-averaged Bloch vectors. Highlighted on each are no loss (solid circles), high loss (open circles), and the fixed point of complete loss (star), where the channel is described by Eq.~\eqref{eq:qubitchannel_maxloss}.
    }
\end{figure*}
When pure, ideal GKP states are damped and then undergo ideal error correction, they are projected onto a different ideal GKP state depending on the syndrome $\vec{m}$.  This process can take Pauli eigenstates to states that lie outside the stabilizer Paulihedron \cite{baragiola_all-gaussian_2019}, a region of the Bloch sphere indicating distillable magic \cite{baragiola_all-gaussian_2019}. This behavior is illustrated in the single-syndrome Bloch vector in Fig.~\ref{fig:damping_noloss}. Interestingly, even on average, error-corrected damped Pauli eigenstates often lie outside the stabilizer Paulihedron. 

For the case of extreme damping, $\beta \rightarrow \infty$, any input state (GKP or otherwise) is damped to the vacuum state, which on average error corrects to a distillable state on the Hadamard axis~\cite{baragiola_all-gaussian_2019}. This channel is identical to the channel for complete loss, $\gamma = 1$, which also converts any input state to the vacuum:
      \begin{align} \label{eq:qubitchannel_maxloss}
        \qubitChannel = \mathcal{E}_\text{EC} \circ \outprod{\text{vac}}{\text{vac}} \circ \projGKP .
    \end{align}
This is a case of the \emph{replacement channel}, here replacing any input GKP state with the error-corrected vacuum state. Averaged over syndromes, the error-corrected vacuum gives the mixed state~\cite{baragiola_all-gaussian_2019}
    \begin{align} \label{eq:avgECvacstate}
        \op \rho = p_+ \outprod{+ \bar H}{+ \bar H} + p_- \outprod{- \bar H}{- \bar H},
    \end{align}
where $\ket{\pm \bar{H}}$ are eigenstates of the Hadamard operator, $p_+\approx 0.91024$ (found numerically), and $p_+ + p_- = 1$. 
A representation of the channel has Kraus operators $\bar{K}_{\pm,j} = \sqrt{p_\pm} \, \outprod{\pm \bar H}{j}$, with $j$ running over any basis. Using the Hadamard eigenstates as this basis with the relation $\bar{Y}\ket{\pm H} = \pm i \ket{\mp \bar{H}}$, we get
    \begin{align}
        \bar{K}_{+,+} &= \frac{\sqrt{p_+}}{2} \left( \bar{I} + \frac{1}{\sqrt{2}}(\bar{X} + \bar{Z} ) \right),
        \\
        \bar{K}_{+,-} &= \frac{\sqrt{p_+}}{2} \left( i \bar{Y} + \frac{1}{\sqrt{2}}( \bar{X} - \bar{Z} ) \right),
        \\
        \bar{K}_{-,-} &= \frac{\sqrt{p_-}}{2} \left( \bar{I} - \frac{1}{\sqrt{2}}(\bar{X} + \bar{Z} ) \right),
        \\
        \bar{K}_{-,+} & = \frac{\sqrt{p_-}}{2}  \left( -i \bar{Y} + \frac{1}{\sqrt{2}}( \bar{X} -\bar{Z} ) \right). 
    \end{align}
This gives a process matrix,
    \begin{align} \label{eq:vacuumEC_channel}
        \mat{\chi} 
        &= \frac{1}{4}
        \begin{pmatrix}
            1 & \frac{2p_+ - 1}{\sqrt{2}} & 0 & \frac{2p_+ - 1}{\sqrt{2}} \\
            \frac{2p_+ - 1}{\sqrt{2}} & 1 & -i\frac{2p_+ - 1}{\sqrt{2}} & 0 \\
            0 & i\frac{2p_+ - 1}{\sqrt{2}} & 1 & -i\frac{2p_+ - 1}{\sqrt{2}} \\
            \frac{2p_+ - 1}{\sqrt{2}} & 0 & i\frac{2p_+ - 1}{\sqrt{2}} & 1 
        \end{pmatrix}. 
    \end{align}
Ideal GKP error correction of the vacuum was shown to produce distillable GKP magic states for almost all syndrome outcomes~\cite{baragiola_all-gaussian_2019}.

\subsection{The pure-loss channel} \label{sec:averaged_loss}

Our primary concern is the pure-loss channel, where the light scattered out of the signal mode is irreparably lost and does not impinge upon a detector. One can proceed with this calculation using the Kraus representations in Sec.~\ref{sec:representations_losschannel}; we find that averaging over the heterodyne outcomes is simpler,
\begin{align}
    \chi_{\paulivec\paulivec'}(\vec{m}) 
     &= 
      \frac{1}{4} 
     \int_{-\infty}^\infty \frac{d^2\mu}{\pi} \, r^\het_{\paulivec}(\mu,\vec{m}) \big[ r^{\het}_{\paulivec'}(\mu,\vec{m}) \big]^*.
\end{align}
Inserting the Bloch-vector components for heterodyne detection, Eq.~\eqref{eq:heterodyne-heralded_Blochvector}, gives an expression involving Gaussian integrals that can be performed analytically; see Appendix~\ref{Appendix:Averaging_loss_outcome} for details. The result is a four-dimensional Siegel $\Theta$ function,
\begin{align} \label{eq:matrixelement_avgloss}
    \chi_{\paulivec\paulivec'}(\vec{m}) 
     & =
      \frac{1}{4} 
     \frac{e^{2\beta}e^{\pi i(\vec{u}_{\paulivec\paulivec'}^{-})^\tp (\mat{I} \otimes \mat{\Omega}) \vec{u}_{\paulivec\paulivec'}^{+}} }{ \pi (c_\theta + e^\beta )^2} 
     \Theta \begin{bmatrix} - (\mat{I} \otimes \mat{\Omega})\vec{u}_{\paulivec\paulivec'}^{+} \\ \vec{u}_{\paulivec\paulivec'}^{-} \end{bmatrix} \left(\vec{0} ; \mat{T}\right),
\end{align}
with $\vec{z} = \vec{0}$ and the parameters
    \begin{align}
        \mat{T} &\coloneqq \mat{I}\otimes \mat{\tau}^\het - \frac {s^2_{\theta} }{(c_\theta + e^\beta)^2} (i\mat{X} \otimes \mat{I} + \mat{\Omega} \otimes \mat{\Omega}),
        \\
        \vec{u}_{\paulivec\paulivec'}^{\pm} &\coloneqq \begin{bmatrix}
        \vec{w}_{\paulivec}^{\pm} \\
        \pm\vec{w}_{\paulivec'}^{\pm}  
        \end{bmatrix} \, ,
\end{align}
where $\mat{\tau}^\het$ and $\vec{w}_{\paulivec}^\pm$ are given in Eqs.~\eqref{eq:params_dampherald}, $\mat{X} \coloneqq \left( \begin{smallmatrix} 0 & 1 \\ 1 & 0 \end{smallmatrix} \right)$, and $\otimes$ refers to the tensor product between matrices. Unlike the heralded cases, the matrix $\mat{T}$ is not diagonal, so the Siegel $\Theta$ function does not split up nicely into a product of Jacobi $\theta$ functions. Nevertheless, the analytic form in Eq.~\eqref{eq:matrixelement_avgloss} can still be evaluated efficiently.

In Fig.~\ref{Fig:processmatrix_loss}, we show the average-loss process matrices for fixed damping $\beta = 0.1$ (corresponding to a $\Delta_\text{dB} =$ 10 dB GKP state) for various loss rates. The top row shows process matrices for the post-selected syndrome $\vec{m} = (0,0)$, and the bottom row shows process matrices averaged over $\vec{m}$. We focus our discussion here on the latter. We observe that compared to the case of damping only (Fig.~\ref{fig:damping_noloss}), the logical effects of loss in the low-loss regime are effective Pauli-X and Pauli-Z errors at the same rate. Pauli-Y errors occur at a reduced rate due to inherent bias in the square-lattice GKP code, requiring larger shifts to induce a logical $Y$ error. As loss increases, non-Pauli noise increases (the magnitudes of the off-diagonal elements in the process matrix become appreciable) and becomes an important driver of qubit dynamics. This is evident in the Bloch-vector curves, which initially move towards the center of the Bloch sphere under the low-loss Pauli noise. At around $\gamma = 0.7$, non-Pauli effects dominate as the channel nears Eq.~\eqref{eq:qubitchannel_maxloss}. At extremely high levels of loss, $\gamma \approx 1$, all states are driven towards the fixed point of complete loss with GKP error correction, a distillable mixture of Hadamard eigenstates.

\subsection{Average channel fidelity}
We quantify logical channels for damping and loss using the average channel fidelity, 
\begin{align}
    F_\text{avg}(\mathcal{E}) 
    := \int d\psi\, \bra{\psi} \mathcal{E}\big(\ketbra{\psi}{\psi} \big) \ket{\psi},
\end{align}
which serves to determine how well a quantum channel $\mathcal{E}$ approximates the identity channel.
In our setting, the average fidelity quantifies how well GKP error correction serves to eliminate the destructive effects of damping and loss on logically encoded information.
For single-qubit CPTP channels, the average fidelity can be evaluated using the formula~\cite{Jones02,Nielsen02}
    \begin{align}
        F_\text{avg}(\mathcal{E}) 
        &= \frac{2 \chi_{0,0} + 1}{3},
    \end{align}
which follows from the fact that the entanglement fidelity is simply the $\chi_{0,0}$ element of the process matrix.
We focus on the syndrome-averaged channels as shown in the second row of Fig.~\ref{fig:damping_noloss} for damping only and of Fig.~\ref{Fig:processmatrix_loss} for damping and loss. Since the logical channels we derive for the loss plus damping channel are not trace preserving, we renormalize them as described in the caption of Fig.~\ref{fig:legend} before using the formula above. 

\begin{figure}[]
    \includegraphics[width=0.42\textwidth]{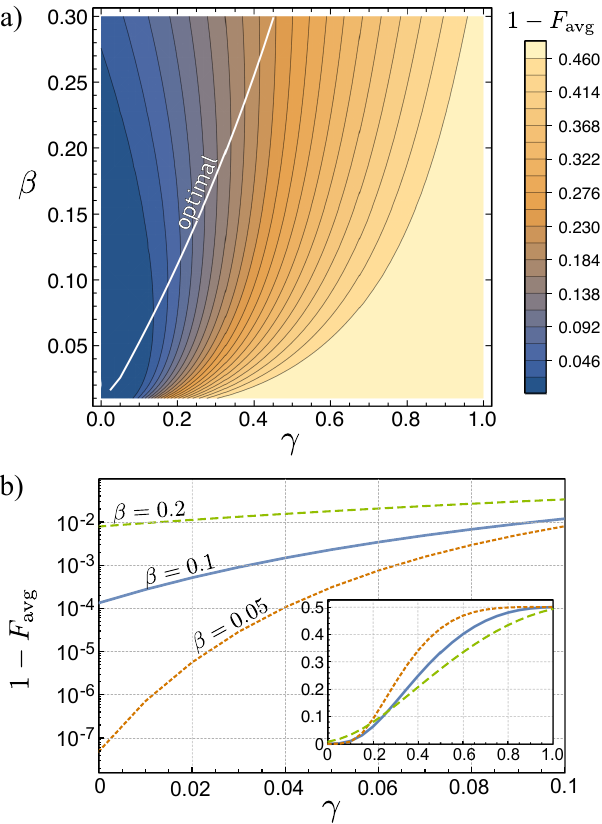}
    \caption[]{\label{Fig:avg_fidelity} 
    (a) Average channel infidelity of the syndrome-averaged logical channel for varying damping and loss rates. optimal $\beta$ for each $\gamma$ is indicated by the white line. (b) Cross sections showing infidelities for several values of $\beta$ for the low-loss regime. The inset figure shows these curves for all loss values.  
    }
\end{figure}

A contour plot of the average channel infidelity, $1-F_\text{avg}$, for varying the damping parameter and loss parameter is shown in Fig.~\ref{Fig:avg_fidelity}(a). Because loss does not act uniformly in Fock space or phase space (see Fig.~\ref{fig:losscartoon} for a cartoon depiction), an optimal value for the damping parameter $\beta$ for each $\gamma$ exists~\cite{hastrup_analysis_2023,shaw_stabilizer_2024}, shown by the white line. This stems from the fact that it can be detrimental to have too many spikes far from the origin in phase space because distant spikes are dragged by loss into measurement bins that correspond to logical errors. The optimal $\beta$ balances the width of each spike with the support of spikes far from the origin to maximize the average fidelity. In Fig.~\ref{Fig:avg_fidelity}(b), we plot cross sections of the contour plot at several values of damping, $\beta = (0.2, 0.1, 0.05)$, corresponding to $\Delta_\text{dB} \approx (7, 10, 13)$ dB of effective squeezing in the input states before loss.
The small infidelities serve as evidence that fault tolerance can be achieved if the loss is small enough, a conclusion borne out in numerical studies that group damping and loss into a single Gaussian random noise channel \cite{fukui2021alloptical} and concatenate with outer qubit codes~\cite{tzitrin2021passive,walshe2025}. Note that similar plots can be found in Refs.~\cite{hastrup_analysis_2023,shaw_stabilizer_2024}. The inset plots the curves for all values of loss, revealing the reversal of the curves due to high-quality input states being damaged more by large amounts of loss, as discussed above.

\begin{figure*}[t]
    \includegraphics[width=0.9\textwidth]{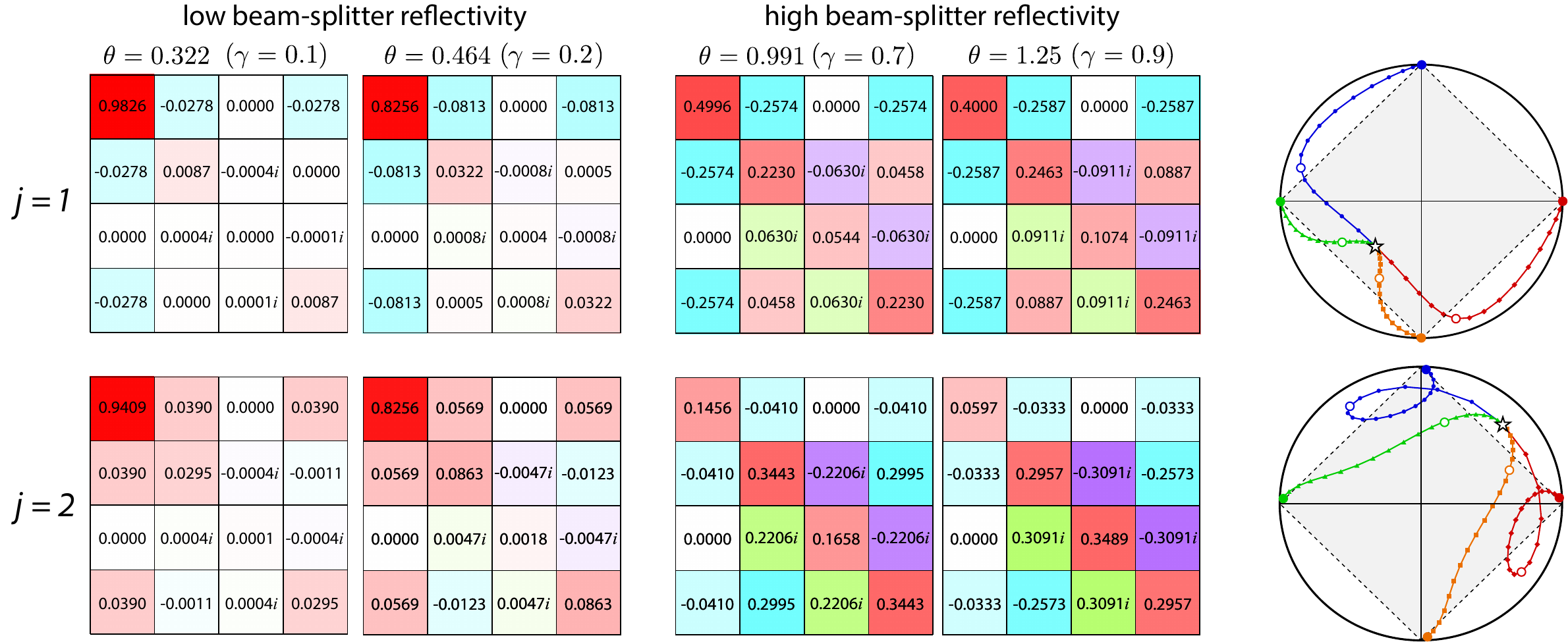}
    \caption[]{\label{fig:Blochsphere_heralded_loss}
    Process matrices for fixed damping ($\beta = 0.1$), followed by loss heralded on photon counting with outcome $j$ and then GKP error correction. To the right are trajectories in the XZ plane of the Bloch sphere for Pauli eigenstates. Each curve shows the Bloch vector as the loss parameter $\gamma$ varies from 0 (no loss) to 1 (complete loss). 
    On each curve, three points are marked: no loss $\gamma = 0$ (solid circle), high loss $\gamma = 0.7$ (open circle), and complete loss $\gamma = 1$ (star). Note that the beam splitter angle is related to the loss parameter via $\theta = \sin^{-1} \sqrt{\gamma}$. Also, for $\gamma = 1$, the probability for odd photon counts vanishes.}
\end{figure*}

\section{Photon subtraction} \label{sec:photon_subtraction}

While our primary concern is the pure-loss channel studied above, the tools we have developed give us access to other potentially useful situations. One such situation is photon subtraction, where photon counting is performed on the loss mode and we herald on non-zero photon counts. 
Photon subtraction is typically performed in the regime of low beam-splitter reflectivity to pick off a small portion of the state in the data mode and implement the ideal subtraction $\op{a}^n$ on the state in that mode. From the Kraus operators in Eq.~\eqref{eq:loss_KrausFock}, we see that the physical evolution does not perform ideal photon subtraction, instead applying $\op{a}^j (1-\gamma)^{\frac{\op{n}}{2}}$, which includes predamping determined by the loss rate (beam-splitter reflectivity).
Nevertheless, when the reflectivity is small enough, $\theta \ll 1$ ($\gamma \ll 1$), this approximates ideal photon subtraction well since $(1-\gamma)^{\frac{\op{n}}{2}} \approx \hat{I}$. For lower transmissivity, a state in the data mode undergoes significant unwanted damping before photons are subtracted.
At the extreme limit of high reflectivity, $\theta = \frac{\pi}{2}$ ($\gamma = 1$), the beam splitter acts to swap in the data and ancilla modes. In this case, the vacuum state proceeds towards error correction, and the input GKP state is sent to the photon counter. 

Subtracting photons from a GKP state followed by GKP error correction can produce highly non-Pauli logical channels. The reason is that photon subtraction generally does not respect the code space and drives an input state out into the error subspaces (the \emph{wilderness} space)~\cite{baragiola_all-gaussian_2019}. Leveraging this fact could be useful for producing magic states when it is applied to GKP Pauli eigenstates. The case of vacuum detection ($j=0$) on the loss mode was already considered above; it is, indeed the, situation where the state experiences only damping, with the total damping determined by $\beta_\text{tot} = \beta -\ln{c_\theta} = \beta + \ln{ (1 -\gamma)^{-\frac{1}{2}}}$. Thus, varying $\gamma$ is equivalent to varying $\beta_\text{tot}$; this case is shown in Fig.~\ref{fig:damping_noloss}.

In Fig.~\ref{fig:Blochsphere_heralded_loss}, we examine the cases where up to two photons are subtracted from the loss mode prior to GKP error correction. In each row, we show syndrome-averaged process matrices for a fixed input damping $\beta = 0.1$. To the right, we plot Bloch-sphere trajectories as a function of the beam-splitter transmissivity through the loss parameter $\gamma = \sin^2 \theta$. For low loss ($\gamma = 0.1$), the logical channels are relatively close to the identity channel, with higher photon counts deviating more than lower photon counts. 

At higher levels of loss, the channels deviate significantly from one another until the reflectivity nears 1. At this point, nearly all of the input GKP state is sent to the photon counter, and vacuum is sent to be GKP error corrected. The limiting channel, when the beam splitter is fully reflective ($\gamma \rightarrow 1$), takes the form of the replacement channel in Eq.~\eqref{eq:vacuumEC_channel}. This case is denoted by the stars in the Bloch-vector plots in Fig.~\ref{fig:Blochsphere_heralded_loss}. In this limit, any input GKP state is projected into a mixed state of the form $\bar \rho^{e/o} = p^{e/o}_+ \outprod{+H}{+H} + p^{e/o}_- \outprod{-H}{-H}$, where $e/o$ indicates that the mixture probabilities depend on the even or odd parity of the heralded outcome, respectively. 
We find that even parity ($j=0,2,\dots$) gives $p^e_+ \approx 0.91024$ , and odd parity ($j=1,\dots$) gives $p^o_+ \approx 0.26822$. Note that the even-parity probability $p^e_+$ is equal to $p_+$ in the expression for the logical state obtained by error-correcting the vacuum [see Eq.~\eqref{eq:avgECvacstate}]. 
This is because in this limit, the damped input GKP state is swapped with the vacuum, so the heralding outcome does not affect the error-corrected state.

For odd-parity outcomes, the limiting logical state arises from error correcting a single-photon state. This can be understood by examining how a Kraus operator for loss with fixed $j$, $\krauslossphoton_j$ [Eq.~\eqref{eq:loss_KrausFock}], acts on an input GKP state. First, note that when $\gamma = 1$, strictly speaking, the probability of odd photon counts vanishes because the GKP code space is even parity; \emph{i.e.,}~it has support only on even Fock states~\cite{grimsmo_quantum_2020}. As $\gamma \to 1^-$, the first effect of $\krauslossphoton_j$ is extreme damping that draws the input state nearly to vacuum, with a tiny but nonzero amplitude on higher Fock states, but still, only the even ones are occupied. Then, $\op{a}^j$ subtracts $j$ photons. When $j$ is odd, this must result in a new state that has only an odd number of photons. In general, this will be a superposition of $\ket 1$, $\ket 3$, $\ket 5$, etc. But in the limit of extreme loss ($\gamma \to 1^-$), all amplitudes limit to zero except that for $\ket 1$, resulting in a single-photon state afterward, regardless of the input state. This state is then sent to be error corrected.

We emphasize that, while there could be practical benefits to engineering non-Pauli channels by counting photons in the loss mode, to assess the viability of such schemes, one would have to account for the downside of nondeterministic implementation. This includes a proper accounting for the probability of each outcome $j$, which depends not only on $j$ but also on the input state.

\section{Discussion}

In this work, we focused on a Steane implementation of ideal GKP error correction using a two-step syndrome readout followed by standard-binning decoding. Our analytical expressions allow for quick numerical and analytical analysis of the error-corrected damping plus loss given an arbitrary loss rate and damping strength. 
We note that optical architectures are more amenable to a different flavor of GKP error correction (EC), called \textit{Knill EC}, in which a noisy input state is teleported through a GKP Bell pair prepared by entangling two ancillae grid states on a beam splitter~\cite{walshe_continuous-variable_2020}. The noise properties of Steane EC and Knill EC differ when noisy ancillae are used~\cite{mahnaz2025, marqversen2025}, but for ideal ancillae,  their respective Kraus operators are identical up to sign changes~\cite{conradthesis, mahnaz2025}. Thus, our results may be employed for Knill EC, too. 

An approach complementary to the calculations here could employ the stabilizer subsystem decomposition (SSSD) of Shaw \emph{et al.}~\cite{shaw_stabilizer_2024}, which divides the CV Hilbert space into that of a GKP qubit and a virtual error subsystem~\cite{pantaleoni_modular_2020, pantaleoni_zak_2023} using the Zak representation. At its heart, the SSSD is founded on ideal GKP error correction with standard binning decoding, and the stabilizer subsystem trace is equivalent to averaging over GKP error correction.

Various other techniques could be used to find other loss-based logical channels for comparison. For example, Jafarzadeh \emph{et al.}~\cite{mahnaz2025} considered a numerically optimized lookup table for finite-energy error correction; a similar technique might improve the quality of logical channels under damping and loss. Another useful comparison would be to pre- or postamplify to convert loss into a random displacement channel before error correction~\cite{albert_performance_2018}. As this procedure adds noise, the logical channels should be farther from the identity channel than those studied here, but their analytic forms should be diagonal Pauli channels~\cite{mahnaz2025}. On the other hand, using a near-optimal recovery via the transpose channel or Petz map \cite{zheng_near-optimal_2024}, it has been shown that GKP codes almost achieve the capacity of the pure-loss channel~\cite{zheng_near-optimal_2024, zheng_performance_2024}. From a logical-channel perspective, optimal recovery should provide a bound on the channel fidelity to which physical techniques, like standard GKP error correction, can be compared and benchmarked.

\begin{acknowledgments}
The authors thank Matthew Stafford, Lucky Antonopoulos, Dominic Lewis, Nicholas Funai, Mahnaz Jafarzadeh, and Rafael Alexander for discussions. We acknowledge support from the Australian Research Council Centre of Excellence for Quantum Computation and Communication Technology (Project No. CE170100012).  T.M.\ was supported by JST, CREST Grant Number JPMJCR23I3, Japan. N.C.M.\ was supported by an ARC Future Fellowship (Project No. FT230100571)
\end{acknowledgments}

\onecolumngrid

\appendix

\section{Representations of the pure loss channel}\label{Appendix:losschannelRep}

Here, we provide derivations for several representations of the pure-loss channel, motivated by number (discrete), heterodyne (continuous), and homodyne (continuous) measurements. The first two are used in the main text to derive logical qubit channels.

\subsection{Photon counting representation}\label{Appendix:losschannelpho}
Consider the standard model of pure loss with the Stinespring dilation
\begin{equation} 
\begin{split} 
    \Qcircuit @C=2.0em @R=2.5em  
    {
    &&&\lstick{(out)}  & \varbs{1} & \rstick{(in)}  \qw \\
    &&&\lstick{\bra{j}}  & \qw & \rstick{\ket{0}} \qw
    } 
\end{split}
\qquad\qquad ,
\end{equation}
where we consider photon counting of the ancillary mode with outcome $j\in\integers_{\geq 0}$. The circuit corresponds to a Kraus operator
\begin{align}
\krauslossphoton_j=(\op{I}_1\otimes\bra{j}_2)\op{B}_{12}(\theta)(\op{I}_1\otimes\ket{0}_{2}),
\end{align} 
where subscripts $1 \text{and} 2$ refer to the top and bottom wires. We represent the identity operators in the Fock basis, $\op{I}=\sum_{n=0}^\infty\ketbra{n}{n}$, and apply the beam splitter to the right, using the relation~\cite{scully_quantum_1997}
\begin{align}
\op{B}_{12}(\theta)(\ketsub{n}{1}\otimes\ketsub{0}{2})
&=
\sum_{k=0}^n
\sqrt{\frac{n!}{k!(n-k)!}}(t_{\theta})^k c_{\theta}^{n}
\ketsub{n-k}{1}\ketsub{k}{2},
\end{align}
to get the expression 
\begin{align}
\krauslossphoton_j
&=
\sum_{n,m,=0}^\infty\sum_{k=0}^n
\sqrt{\frac{n!}{k!(n-k)!}}(t_{\theta})^k c_{\theta}^{n}(\outprodsubsub{m}{m}{1}{1}\otimes\brasub{j}{2})(\outprodsubsub{n-k}{n}{1}{1}\otimes\ketsub{k}{2}).\label{eq.bef_expansion_of_par}
\end{align}
From here, we use $\inprodsubsub{m}{n-k}{1}{1} = \delta_{m,n-k}$ and $\inprodsubsub{j}{k}{2}{2} = \delta_{j,k}$
and the sifting property of Kronecker $\delta's$, $\sum_l \delta_{j,l}a_l=a_j$ \cite{arfken_mathematical_2008}, to replace $k$ with $j$ and  $m=n-j$ and obtain
\begin{align}
\krauslossphoton_j&=\sum_{n=0}^\infty\sqrt{\frac{n!}{j!(n-j)!}}(t_{\theta})^j c_{\theta}^{n}\outprodsubsub{n-j}{n}{1}{1}.
\end{align}
Note that all terms $n<j$ vanish, but we formally include them in the summation. From here, we can use the relation $\op{a}^j\ket{n}=\sqrt{\frac{n!}{(j-n)!}}\ket{n-j}$ \cite{scully_quantum_1997} to obtain
\begin{align}
\krauslossphoton_j
&=\sum_{n=0}^\infty\frac{1}{\sqrt{j!}}(t_{\theta})^j c_{\theta}^{n}\op{a}^j\outprodsubsub{n}{n}{1}{1}.
\end{align}
Finally, using the fact that $c_{\theta}^{n}\ket{n}=(c_{\theta})^{\op{n}}\ket{n}$, the sum over  $n$ gives an identity, and we arrive at the final form of the Kraus operator,
    \begin{align}
        \krauslossphoton_j
            = t^j_\theta \frac{\op{a}^j}{\sqrt{j!}} c^{\op{n}}_\theta.
    \end{align}

\subsection{Heterodyne representation} \label{Appendix:losschannelhet}
An alternative form of the loss channel employs heterodyne detection of the ancillary mode with outcome $\mu \in \complex$. The Stinespring dilation is 
\begin{equation} 
\begin{split}
    \Qcircuit @C=2.0em @R=2.5em  
    {
    &&&\lstick{(out)}     &\varbs{1} &\rstick{(in)}  \qw \\
    &&&\lstick{\bra{\mu}} &\qw       &\rstick{\ket{0}} \qw
    } 
\end{split}
\qquad\qquad ,
\end{equation}
corresponding to Kraus operator
\begin{align}
\krauslosshet(\mu)=(\op{I}_1\otimes\brasub{\mu}{2})\op{B}_{12}(\theta)(\op{I}_1\otimes\ketsub{0}{2}).
\end{align}
We expand the identity operators in the coherent-state basis $\op{I} = \frac{1}{\pi} \int d^2\alpha \outprod{\alpha}{\alpha}$ and apply the beam splitter to the right using the relation~\cite{weedbrook_gaussian_2012}
\begin{align}
\op{B}_{12}(\theta)(\ketsub{\alpha}{2}\otimes\ketsub{\beta}{2}) = \ketsub{\alpha c_{\theta}-\beta s_{\theta}}{1} \otimes \ketsub{\alpha s_{\theta}+\beta c_\theta}{2}
\end{align}
In our case $\beta=0$, which gives
\begin{align}
\krauslosshet(\mu)=\iint_{-\infty}^{\infty}\frac{d^2\alpha d^2\beta}{\pi^2}(\outprodsubsub{\beta}{\beta}{1}{1}\otimes\brasub{\mu}{2})(\outprodsubsub{\alpha c_{\theta}}{\alpha}{1}{1}\otimes\ketsub{\alpha s_{\theta}}{2}).
\end{align}
Evaluating the two inner products $\inprodsubsub{\beta}{\alpha c_{\theta}}{1}{1}$ and $\inprodsubsub{\mu}{\alpha s_{\theta}}{2}{2}$, using the relation $\inprod{\alpha}{\beta}=e^{-\frac{1}{2}(|\alpha|^2+|\beta|^2-2\alpha^*\beta)}$ \cite{walls_quantum_2008}, gives 
\begin{align}
\krauslosshet(\mu)
&=e^{-\frac{1}{2}|\mu|^2}\iint_{-\infty}^{\infty}\frac{d^2\alpha d^2\beta}{\pi^2}e^{-\frac{1}{2}|\alpha |^2+\mu^*\alpha s_{\theta}}e^{-\frac{1}{2}(|\beta|^2-2\beta^*\alpha c_{\theta})}\outprodsubsub{\beta}{\alpha}{1}{1}.
\end{align}
From here, we use the relation $\int_{-\infty}^\infty\frac{d^2\beta}{\pi}e^{-\frac{1}{2}(|\beta|^2-2\beta^*z)}\ket{\beta}=e^{\frac{1}{2}|z|^2}\ket{z}$~\cite{scully_quantum_1997}, with $z = \alpha c_{\theta}$, to get
\begin{align}
\krauslosshet(\mu)
&=
e^{-\frac{1}{2}|\mu|^2}\int_{-\infty}^{\infty}\frac{d^2\alpha}{\pi}e^{-\frac{1}{2}|\alpha |^2(1+c^2_{\theta})+\mu^*\alpha s_{\theta}}\outprodsubsub{\alpha c_{\theta}}{\alpha}{1}{1}.
\end{align}
From here, we use the eigenvalue relation,
$e^{\mu^*\alpha s_{\theta}}\ket{\alpha c_{\theta}}=e^{\mu^*\op{a} t_{\theta}}\ket{\alpha c_{\theta}}$, and the formula $\eta^{\op{n}}\ket{\alpha} = e^{-\frac{|\alpha|^2}{2}\left(1-|\eta|^2\right)}\ket{\eta\alpha}$~ \cite{albert_performance_2018}, with $\eta = c_{\theta}$, to arrive at the final form for the Kraus operator,
\begin{align}
\krauslosshet(\mu) 
        &= e^{-\frac{1}{2}|\mu|^2} e^{\mu^* t_{\theta} \op{a} } c_{\theta}^{\op{n}} .
\end{align}

\subsection{Homodyne representation} \label{Appendix:losschannelhomodyne}
We present the homodyne representation to complete the story for standard quantum-optical measurements. 
Consider an arbitrary quadrature $ \op{p}_\phi \coloneqq \op{R}^\dagger(\phi)\op{p}\op{R}(\phi)=\op{q}\sin{\phi}+\op{p}\cos{\phi} $ with phase-delay operator $\op{R}(\phi) = e^{i\phi\hat{n}}$. A measurement of $\op{p}_\phi$ with outcome $x$ is described by projections onto the left eigenstates $\brasub{x}{p_{\phi}} = \pbra{x} \op{R}(\phi)$. The Stinespring dilation of the loss channel with homodyne measurements is given by the circuit
\begin{equation} 
\begin{split} \label{Cir:loss_circuit_homodyne}
    \Qcircuit @C=2.0em @R=3.0em  
    {
    \lstick{(out)}  & \varbs{1} & \rstick{(in)}  \qw \\
    \lstick{\brasub{x}{p_\phi}}  & \qw & \rstick{\ket{0}} \qw
    } 
    \qquad \quad
    \raisebox{-1.5em}{=}
    \qquad \quad
    \Qcircuit @C=1.2em @R=1.5em  
    {
    &\lstick{(out)} &\gate{R^\dagger(\phi)R(\phi)} &\varbs{1}      &\rstick{(in)} \qw \\
    &\lstick{\brasub{x}{p}}   &\gate{R(\phi)} &\qw       &\rstick{\ket{0}} \qw
   }  
    \qquad\quad
    \raisebox{-1.5em}{=}
     \qquad
    \Qcircuit @C=1.5em @R=2.5em  
    {
    &\lstick{(out)} &\gate{R^\dagger(\phi)} &\varbs{1} &\gate{R(\phi)}     &\rstick{(in)} \qw \\
    &                &\lstick{\brasub{x}{p}}   &\qw       &\rstick{\ket{0}} \qw
   } 
\end{split}
\qquad.
\end{equation}
We inserted $\op{I} = \op{R}^\dagger(\phi) \op{R}(\phi)$ and then used the fact that identical Gaussian operations commute with a beam splitter to move a phase delay on each mode to the right. The phase delay on the bottom mode vanishes because the vacuum is invariant under rotations.
The Kraus operator for the circuit is
    \begin{align} \label{appeq:homodynekraus}
    \krauslossqadrot(x)
        \coloneqq
        \op{R}^\dagger(\phi)(\hat{I} \otimes \pbra{x})\hat{B}_{12}(\theta)(\hat{I}\otimes\ket*{0}) \op{R}(\phi)
        =
        \op{R}^\dagger(\phi) \krauslossqad(x)  \op{R}(\phi),
    \end{align}
 where $\qadrot$ indicates homodyne measurement of the rotated momentum quadrature. 
This circuit indicates that we need to evaluate only the Kraus operator for momentum measurements $\op{L}^\qad(x)$; other quadratures are obtained by applying a phase delay.

We now derive a closed form for $\op{L}^\qad(x)$. First, we follow the same steps as we did for photon counting up to Eq.~\eqref{eq.bef_expansion_of_par}. At this point in the homodyne setting, we encounter an inner product between a momentum eigenstate and a number state, interpreted as a momentum wave function given by~\cite{zettili_quantum_2009} 
    \begin{align}
    \inprodsubsub{x}{k}{p_2}{2}=\frac{1}{\pi^{\frac{1}{4}}\sqrt{2^k k!}}H_{k}(x)e^{-\frac{1}{2}x^2},
    \end{align}
where $H_{k}(x)$ are $k$th order Hermite polynomials~\cite{kim_note_2016}.
Using the above relation, following from Eq.~\eqref{eq.bef_expansion_of_par}, we obtain
    \begin{align}
        \krauslossqad(x) 
        =\frac{1}{\pi^{\frac{1}{4}}}e^{-\frac{1}{2}x^2}\sum_{n=0}^\infty\sum_{k=0}^n \sqrt{\frac{n!}{(n-k)!}}\frac{(2^{-\frac{1}{2}}t_{\theta})^k }{k!}H_{k}(x)c_{\theta}^{n}\outprodsubsub{n-k}{n}{1}{1}.
    \end{align}
We again pull out $\op{a}^k$ and use $\sum_n c_\theta^n \ket{n}\bra{n} = c_{\theta}^{\op{n}}$ to get
    \begin{align}
    \op{L}^{p}(x) 
        =\frac{1}{\pi^{\frac{1}{4}}}e^{-\frac{1}{2}x^2}\sum_{k=0}^\infty \frac{(2^{-\frac{1}{2}}t_{\theta}\op{a})^k }{k!}H_{k}(x)c_{\theta}^{\op{n}}.
    \end{align} 
We use the generating function for Hermite polynomials, $e^{2xt-t^2}=\sum_{n=0}^\infty\frac{H_{n}(x)}{n!}t^n$ \cite{kim_note_2016}, with $t=2^{-1/2}t_{\theta}\op{a}$, to rewrite the expression as
\begin{align}
\krauslossqad(x)
        =\frac{1}{\pi^{\frac{1}{4}}}e^{-\frac{1}{2}x^2}e^{\sqrt{2}x t_{\theta}\op{a}-\frac{1}{2}(t_{\theta}\op{a})^2} c_{\theta}^{\op{n}}.
\end{align}

From Eq.~\eqref{appeq:homodynekraus}, the Kraus operator for arbitrary quadrature measurement is found by conjugating $\krauslossqad(x) $ with phase delays. Using $\op{R}^\dagger(\phi) \op a \op{R}(\phi) = e^{-i\phi} \op{a}$, this gives the final expression for the Kraus operator, 
\begin{align}
\krauslossqadrot(x) 
        =\frac{1}{\pi^{\frac{1}{4}}}e^{-\frac{1}{2}(x-e^{-i\phi}t_{\theta}\op{a})^2}c_{\theta}^{\op{n}}.
\end{align}

\section{Properties of Siegel $\Theta$ functions}\label{Appendix.2.9theta_function}

A $d$-dimensional Siegel $\Theta$ function is defined as \cite{deconinck_computing_2002}
    \begin{align}
        \Theta(\vec{z};\mat{\tau}) 
        \coloneqq 
        \sum_{ \vec n \in\mathbb{Z}^d} e^{2\pi i\left(\frac{1}{2}\vec{n}^\tp\mat{\tau}\vec{n}+\vec{n}^\tp \vec{z}\right)},\label{eq.theta_form}
    \end{align}
where $(z\in\mathbb{C}^d)$ and $\mat{\tau}$ exist in the Siegel upper half-space, the set of symmetric square matrices whose imaginary parts are positive definite.  
Including characteristic vectors $(\vec{v}_1,\vec{v}_2)\in\mathbb{Q}^d$ gives a Siegel function with following characteristics:
    \begin{align}
        \Theta\begin{bmatrix}
        \vec{v}_1\\
        \vec{v}_2
        \end{bmatrix}(\vec{z};\mat{\tau}) 
        \coloneqq
        \sum_{\vec{n}\in\mathbb{Z}^d}e^{2\pi i \left(\frac{1}{2}(\vec{n}+\vec{v}_1)^\tp\mat{\tau}(\vec{n}+\vec{v}_1)+(\vec{n}+\vec{v}_1)^\tp(\vec{z}+\vec{v}_2)\right)}
        =
        e^{2\pi i \left(\frac{1}{2}\vec{v}^{\tp}_1\mat{\tau}\vec{v}_1+\vec{v}^{\tp}_1(\vec{z}+\vec{v}_2)\right)}\Theta(\vec{z}+\mat{\tau}\vec{v}_1+\vec{v}_2;\mat{\tau})
        .\label{eq.2.9_theta_charac_formula}
\end{align}

If we take the complex conjugate of a Siegel $\Theta$ function, it will become 
\begin{align}\label{eq:complex_con_theta}
\left\{\Theta\begin{bmatrix}
        \vec{v}_1\\
        \vec{v}_2
        \end{bmatrix}(\vec{z};\mat{\tau})\right\}^*=\sum_{\vec{n}\in\mathbb{Z}^d}e^{2\pi i \left(\frac{1}{2}(\vec{n}+\vec{v}_1)^\tp(-\mat{\tau})^*(\vec{n}+\vec{v}_1)+(\vec{n}+\vec{v}_1)^\tp(-\vec{z}^*-\vec{v}_2)\right)}=\Theta\begin{bmatrix}
        \vec{v}_1\\
        -\vec{v}_2
        \end{bmatrix}(-\vec{z}^*;-\mat{\tau}^*).
\end{align}
Siegel $\Theta$ functions satisfy a variety of useful properties under modular transformations~\cite{mumford_tata_1983,amir-khosravi_theta_2019}, one of which we call "flip $\Theta$",
    \begin{align}
        \Theta
        \begin{bmatrix}
        \vec{v}_1\\
        \vec{v}_2
        \end{bmatrix}
        (\vec{z};\mat{\tau}) = \sqrt{\det(i \mat{\tau}^{-1})}\,\Theta
        \begin{bmatrix}
        -\vec{v}_2\\
        \vec{v}_1
        \end{bmatrix}
        (\mat{\tau}^{-1}\vec{z};-\mat{\tau}^{-1}) \;e^{2\pi i \left(\frac{1}{2}\vec{z}^\tp(-\mat{\tau}^{-1})\vec{z}+\vec{v}_1^\tp\vec{v}_2\right)}.\label{eq.2.9_flip_theta_characteristics}
\end{align}
A product of Siegel $\Theta$ functions is another Siegel $\Theta$ function (of larger dimension) with block-diagonal $\tau$:
    \begin{align}
        \Theta\begin{bmatrix}
        \vec{v}_{1}\\
        \vec{v}_{2}
        \end{bmatrix}
        \left(\vec{z};\mat{\tau}\right)\;\Theta
        \begin{bmatrix}
        \vec{v}'_{1}\\
        \vec{v}'_{2}
        \end{bmatrix}
        \left(\vec{z}';\mat{\tau}'\right)
        =
        \Theta
        \begin{bmatrix}
        \vec{u}_{1}\\
        \vec{u}_{2}
        \end{bmatrix}
        \left(\vec{y};\mat{\tau} \oplus \mat{\tau}' \right),\label{eq:stacking_theta}
    \end{align}
where $\vec{u}_j = (\vec{v}_{j}, \vec{v}'_{j})^\tp$ and $\vec{y} = (\vec{z},\vec{z}')^\tp$.
This relation also applies in reverse: A Siegel $\Theta$ function with block-diagonal $\mat{\tau}$ can be decomposed into a product of Siegel $\Theta$ functions. 
Integrals of Siegel $\Theta$ functions with Gaussian kernels can be evaluated analytically. %
For a $\complex^{n\times m}$ matrix $\mat{A}$, consider the integral
\begin{align}
    \int_{-\infty}^{\infty}  d^n x ~\Theta\begin{bmatrix}
    \vec{v}_1\\
    \vec{v}_2
    \end{bmatrix}(\mat{A}\vec{x};\mat{\tau}) \,e^{-\frac{1}{2}\vec{x}^\tp \mat{\Sigma} \vec{x}+\vec{\nu}^\tp \vec{x}}
    =
    \sum_{\vec{n}\in\mathbb{Z}^m}  e^{2\pi i \left(\frac{1}{2}(\vec{n}+\vec{v}_1)^\tp\mat{\tau}(\vec{n}+\vec{v}_1)+(\vec{n}+\vec{v}_1)^\tp(\vec{v}_2)\right)}
    \int_{-\infty}^{\infty} d^n\vec{x}~  e^{-\frac{1}{2}\vec{x}^\tp \mat{\Sigma} \vec{x}+\vec{\xi}^\tp \vec{x}},
\end{align}
where $\vec{x}$ and $\vec{\nu}$ are $\reals^n$ vectors, $\mat \Sigma$ is the $\complex^{m\times m}$ covariance matrix, $\vec{v}_1$ and $\vec{v}_2$ are $\rationals^{m}$ characteristic vectors, and $\mat{\tau}$ is an $\complex^{m\times m}$ matrix in the Siegel upper half plane. To arrive at the right-hand side, we expanded the Siegel $\Theta$ function, collected terms that depend on the integration variables, and introduced the shorthand notation $\vec{\xi} \coloneqq 2\pi i\mat{A}^\tp(\vec{n}+\vec{v}_1)+\vec{\nu}$.
The remaining multivariate Gaussian integral evaluates to 
\begin{align}
    \int_{-\infty}^{\infty} d^{ n}\vec{x}\,e^{-\frac{1}{2}\vec{x}^\tp \mat{\Sigma} \vec{x}+\vec{\xi}^\tp \vec{x}}
    &=
    \sqrt{\frac{(2\pi)^{ n}}{\det\mat{\Sigma}}}e^{\frac{1}{2}\vec{\xi}^\tp \mat{\Sigma}^{-1} \vec{\xi}}  .
    \label{eq:multivari_gaussian_int}
\end{align}
Combining terms, expanding out $\vec{\xi}$, and collecting all the summation variables, we can write out our new theta and Gaussian functions:
\begin{align}
    \int_{-\infty}^{\infty}  d^n x ~\Theta\begin{bmatrix}
    \vec{v}_1\\
    \vec{v}_2
    \end{bmatrix}(\mat{A}\vec{x};\mat{\tau})e^{-\frac{1}{2}\vec{x}^\tp \mat{\Sigma} \vec{x}+\vec{\nu}^\tp \vec{x}}
    &=\sqrt{\frac{(2\pi)^{ n}}{\det\mat{\Sigma}}}\Theta\begin{bmatrix}
    \vec{v}_1\\
    \vec{v}_2
    \end{bmatrix}(\mat{A}\mat{ \Sigma}^{-1}\vec{\nu};\mat{\tau}+2\pi i\mat{A}\mat{\Sigma}^{-1} \mat{A}^\tp)\, e^{\frac{1}{2}\vec{\nu}^\tp \mat{\Sigma}^{-1} \vec{\nu}}.\label{eq.2.9_theta_gaussian_formula}
\end{align}

\section{Process matrix derivation}\label{Appendix:ProcessMatrixDerivation}
To derive the qubit map, we first need to write down the Kraus operator [Eq.~\eqref{eq:hetecKraus}] using definitions for the Kraus operators of the GKP error correction [Eq.~\eqref{eq:Krausop_GKP_correction}], loss [Eq.~\eqref{eq:loss_KrausCoherent}], and damping $e^{-\beta\op{n}}$,
\begin{align} \label{eq:Krausappendix}
    \qubitKraushet(\mu,\vec{m})
    &=
    e^{-\frac{1}{2}|\mu|^2}\projGKP\op{Z}^\dagger(\{m_p\})\op{X}^\dagger(\{m_q\})e^{t_{\theta}\mu^*\op{a}}e^{-d\op{n}}\projGKP,
\end{align}
where we have combined the damping parameter with the damping portion of loss using $d = -\ln{c_\theta} + \beta$. We first project the Kraus operator
into the GKP subspace by taking the trace with the GKP subspace Paulis to find its Bloch four-vector, Eq.~\eqref{eq:Blochvectorcomponent},
\begin{align}
    r^\het_{\paulivec}(\mu,\vec{m}) 
    &= \frac{1}{\sqrt{\pi}}e^{-\frac{1}{2}|\mu|^2} \sum_{\vec{n}\in\mathbb{Z}^2} \Tr \left[\op{Z}^\dagger(\{m_p\})\op{X}^\dagger(\{m_q\})e^{t_{\theta}\mu^*\op{a}}e^{-d\op{n}}\op{D}\left((2\vec{n}+\paulivec)\sqrt{\frac{\pi}{2}}\right)
    e^{i\pi \vec{n}^\tp \mat{\Omega} \paulivec}\right].
\end{align}
We expressed the GKP subspace Pauli operator as a sum of displacements using Eq.~\eqref{eq:Pauli_GKP_Subspace}. 
Cycling the operators and performing the trace in the coherent-state basis gives 
\begin{align}
    r^\het_{\paulivec}(\mu,\vec{m})&=\frac{1}{\sqrt{\pi}}e^{-\frac{1}{2}|\mu|^2}\sum_{\vec{n}\in\mathbb{Z}^2}e^{i\pi \vec{n}^\tp \mat{\Omega} \paulivec}\int_{-\infty}^{\infty} \frac{d^2\alpha}{\pi}\bra{\alpha}e^{-d\op{n}}\op{D}\left((2\vec{n}+\paulivec)\sqrt{\frac{\pi}{2}}\right)\op{Z}^\dagger(\{m_p\})\op{X}^\dagger(\{m_q\})e^{t_{\theta}\mu^*\op{a}}\ket{\alpha}
    \\
    &=\frac{1}{\sqrt{\pi}}e^{-\frac{1}{2}|\mu|^2}\sum_{\vec{n}\in\mathbb{Z}^2}e^{i\pi \vec{n}^\tp \mat{\Omega} \paulivec}\int_{-\infty}^{\infty} \frac{d^2\alpha}{\pi}e^{t_{\theta}\mu^*\alpha}e^{-\frac{1}{2}|\alpha|^2(1-e^{-2d})}\bra{e^{-d}\alpha}\op{D}\left((2\vec{n}+\paulivec)\sqrt{\frac{\pi}{2}}\right)\op{Z}^\dagger(\{m_p\})\op{X}^\dagger(\{m_q\})\ket{\alpha}. \label{appeq:Blochvecderiv1}
\end{align}
In the second line, we evaluated  $e^{t_{\theta}\mu^*\op{a}}$ and $e^{-d\op{n}}$ acting on coherent states; for the latter we used the formula $\eta^{\op{n}}\ket{\alpha} = e^{-\frac{|\alpha|^2}{2}\left(1-|\eta|^2\right)}\ket{\eta\alpha}$~ \cite{albert_performance_2018}.
To complete the evaluation,  we combine the displacements
\begin{align}
    \op{D}\left((2\vec{n}+\paulivec)\sqrt{\frac{\pi}{2}}\right)
     \op{Z}^\dagger(\{m_p\})\op{X}^\dagger(\{m_q\})
    \circeq
    \op{D}\left( \frac{1}{\sqrt{2}} \big[ (2\vec{n}+\paulivec)\sqrt{\pi}-\{ \vec{m} \} \big] \right)
    e^{\frac{i\sqrt{\pi}}{2}(2\vec{n}+\paulivec)^\tp \mat \Omega\{ \vec{m} \}},
\end{align}
where $\{ \vec{m} \} \coloneqq ( \{m_q\}, \{m_p\} )^\tp$, $\mat \Omega$ is the symplectic form in Eq.~\eqref{eq:symplectic-form}, and the equality $\circeq$ is up to a global phase (in this case $e^{i\frac{\{m_q\}\{m_p\}}{2}}$).
With this and the relation \cite{walls_quantum_2008} $\bra{\beta}\op{D}(\gamma)\ket{\alpha} = e^{-\frac{1}{2}\left(|\beta|^2+|\alpha+\gamma|^2-2\beta^*(\alpha+\gamma)\right)} e^{\frac{1}{2}(\gamma\alpha^*-\gamma^*\alpha)}$, we evaluate the inner product in Eq.~\eqref{appeq:Blochvecderiv1} to get
\begin{align}
    r^\het_{\paulivec}(\mu,\vec{m})
    &\circeq \frac{1}{\sqrt{\pi}}
    e^{-\frac{1}{2}|\mu|^2}
    \sum_{\vec{n}\in\mathbb{Z}^2}
    e^{i\pi \vec{n}^\tp \mat{\Omega} \paulivec}
    e^{\frac{i\sqrt{\pi}}{2}(2\vec{n}+\paulivec)^\tp \mat \Omega\{ \vec{m} \}}
    \int_{-\infty}^{\infty}  \frac{d^2\alpha}{\pi}
    e^{ t_{\theta}\mu^*\alpha}e^{-\frac{1}{2}|\alpha|^2(1-e^{-2d})}
    \nonumber
    \\
    &\qquad\times 
    e^{-\frac{1}{2}\left(|e^{-d}\alpha|^2+\left|\alpha+\frac{1}{\sqrt{2}}\eta\right|^2-2(e^{-d}\alpha)^*\left(\alpha+\frac{1}{\sqrt{2}}\eta \right)\right)}e^{\frac{1}{2\sqrt{2}}(\eta\alpha^*-\eta^*\alpha)},
\end{align}
where, for convenience, we define the shorthand notation $\eta =(2n_1+a_1)\sqrt{\pi}-\{m_q\}+i((2n_2+a_2)\sqrt{\pi}-\{m_p\})$.

Now we evaluate the integral over $\alpha = \alpha_R + i \alpha_I$. First, we expand and collect all integration terms,
\begin{align}
    r^\het_{\paulivec}(\mu,\vec{m})
    &\circeq
    \frac{1}{\sqrt{\pi}}
    e^{-\frac{1}{2}|\mu|^2}
    \sum_{\vec{n}\in\mathbb{Z}^2}
    e^{i\pi \vec{n}^\tp \mat{\Omega} \paulivec}
    e^{\frac{i\sqrt{\pi}}{2}(2\vec{n}+\paulivec)^\tp \mat{\Omega}\{ \vec{m} \}}
    e^{-\frac{1}{4}|\eta|^2}
    \int_{-\infty}^{\infty} \frac{d^2\alpha}{\pi}e^{-\frac{1}{2}\vec{\alpha}^\tp \mat{A} \vec{\alpha} + (\mat{B}(\vec{n}+\vec{l}))^\tp\vec{\alpha}},
\end{align}
where $\vec{\alpha} = (\alpha_R, \alpha_I)^\tp$, and
\begin{align}
    \mat{A}= 2(1-e^{-d})\mat{I},\quad \quad
    \vec{l}=-\frac{t_{\theta}\mu^*}{2\sqrt{2\pi}}
    \begin{bmatrix}
    1\\
    i
    \end{bmatrix}+\frac{1}{2}\vec{a}-\frac{1}{2\sqrt{\pi}}\{ \vec{m} \}, \quad \quad
    \mat{B}= -\sqrt{2\pi}
    \begin{bmatrix}
    1 - e^{-d} & -i(1+e^{-d})\\
    i(1+e^{-d}) & 1 - e^{-d}
    \end{bmatrix}.
\end{align}
Using the formula in Eq.~\eqref{eq:multivari_gaussian_int}, we evaluate the Gaussian integral over $\alpha$, substituting in $\eta$, collect all the $\vec{n}$ terms, and recognize that $\mat{B}^\tp \mat{A}^{-1} \mat{B} = \frac{4 \pi}{1 - e^d} \mat{I}$,
\begin{align}
    r^\het_{\paulivec}(\mu,\vec{m})
    &\circeq
    \frac{2 e^{-\frac{1}{2}|\mu|^2}}{\sqrt{\pi|\det\mat{A}|}}
    e^{\frac{\sqrt{\pi}}{2}\paulivec^\tp(\mat{I}+i\mat{\Omega})\{ \vec{m} \}- \frac{1}{4}| \{ \vec{m} \}|^2 -\frac{\pi}{4} |\paulivec|^2 +\frac{2 \pi}{1 - e^d} |\vec{l}|^2}\sum_{\vec{n}\in\mathbb{Z}^2}
    e^{\frac{1}{2} \left(\frac{4 \pi}{1 - e^d} - 2\pi\right) \vec{n}^2 + \vec{n}^\tp \left(\frac{4 \pi}{1 - e^d}\vec{l}+\sqrt{\pi}(\mat{I}+i\mat{\Omega})\{ \vec{m} \}-\pi(\mat{I}-i\mat{\Omega})\paulivec\right)}.
\end{align}
To rewrite the expression in the sum as a Siegel $\Theta$ function of the form in Eq.~\eqref{eq.theta_form}, we factor out $2\pi i$ from the Gaussian and the linear terms involving $\vec{n}$, obtaining the $\mat\tau$ matrix,
\begin{align}
\mat{\tau}=\frac{-i}{2\pi}\left(\frac{4 \pi}{1 - e^d} -2\pi\right)\mat{I}
=i\coth{\frac{d}{2}}\mat{I}.
\end{align} 
With this, we can simplify the linear term by splitting it into parts, 
\begin{align}
\frac{-i}{2\pi}\left(\frac{4\pi}{1-e^d}\vec{l}+\sqrt{\pi}(\mat{I}+i\mat{\Omega})\{ \vec{m} \}-\pi(\mat{I}-i\mat{\Omega})\paulivec\right)
=
\vec{z}+\mat{\tau}\vec{w}_{\paulivec}^{-}+\mat{\Omega}\vec{w}_{\paulivec}^{+}.\label{eq.2.9_vec_z}
\end{align}
where
\begin{align}
\vec{z}&\coloneqq\frac{t_\theta}{\sqrt{2\pi}(e^d-1)}\begin{bmatrix}
-i\\
1
\end{bmatrix}\mu^*,\qquad
\vec{w}_{\paulivec}^{\pm}\coloneqq\frac{1}{2}\left(\paulivec\pm\frac{\{ \vec{m} \}}{\sqrt{\pi}}\right).%
\end{align}
Therefore,
\begin{align} \label{appeq:rvecmidway}
    r^\het_{\paulivec}(\mu,\vec{m})
    &\circeq
    \frac{2}{\sqrt{\pi|\det\mat{A}|}}
    e^{-\frac{1}{2}|\mu|^2}e^{\frac{\sqrt{\pi}}{2}\paulivec^\tp(\mat{I}+i\mat{\Omega})\{ \vec{m} \}-\frac{1}{4}|\{ \vec{m} \}|^2-\frac{\pi}{4}|\paulivec|^2+\frac{2 \pi}{1 - e^d} |\vec{l}|^2}
    \sum_{\vec{n}\in\mathbb{Z}^2}
    e^{2\pi i(\frac{1}{2}\vec{n}^\tp\mat{\tau}\vec{n}+\vec{n}^\tp(\vec{z}+\mat{\tau}\vec{w}_{\paulivec}^{-}+\mat{\Omega}\vec{w}_{\paulivec}^{+}))} 
    .
\end{align}
We could stop here and write out our $\Theta$ function, but we can go further and use the Siegel $\Theta$ with characteristics in Eq.~\eqref{eq.2.9_theta_charac_formula}. To do this, we simplify further by manipulating the exponential terms outside of the sum.  
First, we rewrite the vector $\vec{l}$ in terms of the characteristic defined above,
\begin{align}
    \vec{l}=-\frac{t_{\theta}}{2\sqrt{2\pi}}\mu^*\begin{bmatrix}
    1\\
    i \end{bmatrix}+\vec{w}_{\paulivec}^{-} \label{eq.2.9_new_vec_l},
\end{align}
and then do the same for the other terms in the exponential. This gives
\begin{align}
    e^{\frac{\sqrt{\pi}}{2}\paulivec^\tp(\mat{I}+i\mat{\Omega})\{ \vec{m} \}-\frac{1}{4}|\{ \vec{m} \}|^2-\frac{\pi}{4}|\paulivec|^2+\frac{2 \pi}{1 - e^d} |\vec{l}|^2}
&= e^{\pi i(\vec{w}^{-}_{\paulivec})^\tp\mat\tau\vec{w}_{\paulivec}^{-} -\pi i (\vec{w}^{-}_{\paulivec})^\tp\mat\Omega\vec{w}_{\paulivec}^{+} +2\pi i(\vec{w}^{-}_{\paulivec})^\tp\left(\vec{z}+\mat\Omega\vec{w}_{\paulivec}^{+}\right)}.
\end{align}
Using this expression in Eq.~\eqref{appeq:rvecmidway} and combining terms gives the Bloch-vector components in terms of a Siegel function with characteristics, in Eq.~\eqref{eq.2.9_theta_charac_formula},
\begin{align}
 r^\het_{\paulivec}(\mu,\vec{m})
&\circeq\frac{2 e^{-\frac{1}{2}|\mu|^2}e^{-\pi i(\vec{w}^{-}_{\paulivec})^\tp\mat\Omega\vec{w}_{\paulivec}^{+}}}{\sqrt{\pi|\det\mat{A}|}}\sum_{\vec{n}\in\mathbb{Z}^2}
    e^{2\pi i(\frac{1}{2}(\vec{n}+\vec{w}^{-}_{\paulivec})^\tp\mat{\tau}(\vec{n}+\vec{w}^{-}_{\paulivec})+(\vec{n}+\vec{w}^{-}_{\paulivec})^\tp(\vec{z}+\mat{\Omega}\vec{w}_{\paulivec}^{+}))} 
    \nonumber\\
    &=\frac{2e^{-\frac{1}{2}|\mu|^2}
e^{-\pi i(\vec{w}^{-}_{\paulivec})^\tp \mat{\Omega} \vec{w}_{\paulivec}^{+}}}{\sqrt{\pi|\det\mat{A}|}}
\Theta\begin{bmatrix}
\vec{w}_{\paulivec}^{-}\\
\mat\Omega\vec{w}_{\paulivec}^{+}
\end{bmatrix}\left(\vec{z};\mat{\tau}\right).
\end{align}
The $\mat{\tau}$ matrix of this $\Theta$ function is diagonal with elements of $\coth{\frac{d}{2}}$. When the parameter $d = -\ln{c_\theta} + \beta$ is small, $d \ll 1$, which is the case for low damping and low loss, $\coth \frac{d}{2}$ becomes extremely large. In this case, it may be useful (numerically) to use the flip $\Theta$ Jacobi transformation defined in Eq.~\eqref{eq.2.9_flip_theta_characteristics} to rewrite the Bloch four-vector as:
\begin{align}
r^\het_{\paulivec}(\mu,\vec{m})
&\circeq\frac{ e^\beta e^{-\frac{1}{2}|\mu|^2} e^{\pi i (\vec{w}_{\paulivec}^{-})^\tp \mat{\Omega} \vec{w}_{\paulivec}^{+} } }{ \sqrt{\pi} ( c_\theta + e^\beta )}\Theta\begin{bmatrix}
-\mat{\Omega}\vec{w}_{\paulivec}^{+}\\
\vec{w}_{\paulivec}^{-}
\end{bmatrix}\left( \vec{z}^\het ;\mat{\tau}^\het \right).\label{eq.2.9_c-coefficient_coherent_pauli_bass}
\end{align}
where,
    \begin{align}
    \mat{\tau}^\het \coloneqq  -\mat{\tau}^{-1} =
    i\tanh{\frac{d}{2}}\mat{I},\qquad 
    \vec{z}^\het \coloneqq \mat{\tau}^{-1}\vec{z} = -\frac{s_{\theta}\mu^*}{\sqrt{2\pi}(e^\beta+c_\theta)}
    \begin{bmatrix}
1\\
i
    \end{bmatrix}.
    \end{align}
The $\mat{\tau}^\het$ matrix is diagonal with elements $\tanh{\frac{d}{2}}$ that become \emph{smaller} as $d \ll 1$; this, in theory, will make numerical calculations for low damping and loss more efficient.
\section{Pure loss channel}\label{Appendix:Averaging_loss_outcome}

Using the final analytical form of the post-selected logical loss channel four-vector, Eq.~\eqref{eq.2.9_c-coefficient_coherent_pauli_bass}, we average over the heterodyne detection outcomes to find the process matrix for the pure-loss channel,  
\begin{align}
    \chi_{\paulivec\paulivec'}(\vec{m})
    &=
     \frac{1}{4} 
    \int_{-\infty}^\infty\frac{d^2\mu}{\pi}r^\het_{\paulivec}(\mu,\vec{m}) \big[r^\het_{\paulivec}(\mu,\vec{m}) \big]^*
    \\
    &=
     \frac{1}{4} 
    \frac{e^{2\beta}e^{\pi i [(\vec{w}_{\paulivec}^{-})^\tp \mat{\Omega} \vec{w}_{\paulivec}^{+}- (\vec{w}_{\paulivec'}^{-})^\tp \mat{\Omega} \vec{w}_{\paulivec'}^{+} ]}}{\pi ( c_\theta + e^\beta )^2}
    \int_{-\infty}^\infty\frac{d^2\mu}{\pi}e^{-|\mu|^2} 
    \Theta\begin{bmatrix}
-\mat\Omega\vec{w}_{\paulivec}^{+}\\
\vec{w}_{\paulivec}^{-}
\end{bmatrix}
\left(\vec{z}^\het;\mat{\tau}^\het\right)\left\{\Theta\begin{bmatrix}
-\mat\Omega\vec{w}_{\paulivec'}^{+}\\
\vec{w}_{\paulivec'}^{-}
\end{bmatrix}\left(\vec{z}^\het;\mat{\tau}^\het\right)\right\}^*
\end{align}
In the $\theta$ functions, the integration variable $\mu$ is contained only in $\vec{z}^\het$. 

We now focus on the integral, noting that $\vec{z}^\het$ is a function of the integration variable $\mu$. First, we use Eq.~\eqref{eq:stacking_theta} to rewrite the product of $\theta$ functions as a single $\Theta$ function,
\begin{align}
\Theta\begin{bmatrix}
-\mat{\Omega}\vec{w}_{\paulivec}^{+}\\
\vec{w}_{\paulivec}^{-}
\end{bmatrix}\left(\vec{z}^\het;\mat{\tau}^\het\right)\left\{\Theta\begin{bmatrix}
-\mat{\Omega}\vec{w}_{\paulivec'}^{+}\\
\vec{w}_{\paulivec'}^{-}
\end{bmatrix}\left(\vec{z}^\het;\mat{\tau}^\het\right)\right\}^*
&=
\Theta\begin{bmatrix}
-(\mat{I}\otimes\mat{\Omega})\vec{u}_{\paulivec\paulivec'}^{+}\\
\vec{u}_{\paulivec\paulivec'}^{-}
\end{bmatrix}\left(\vec{y};\mat{I}\otimes\mat{\tau}^\het\right),
\end{align}
where the effect of the complex conjugate on $\theta$ functions is given by Eq.~\eqref{eq:complex_con_theta} and 
\begin{align}
\vec{u}_{\paulivec\paulivec'}^{\pm} \coloneqq 
\begin{bmatrix}
\vec{w}_{\paulivec}^{\pm}\\
\pm\vec{w}_{\paulivec'}^{\pm}
\end{bmatrix},\qquad
\vec{y} \coloneqq 
\begin{bmatrix}
\vec{z}^\het \\
-(\vec{z}^\het )^*
\end{bmatrix}
.
\end{align}
To evaluate the integral using Eq.~\eqref{eq.2.9_theta_gaussian_formula}, we rewrite $\vec{y}$ as  $\vec{y} = \mat{V}\vec{\mu}$, where
\begin{align}
\mat{V} \coloneqq -\frac{s_{\theta}}{\sqrt{2\pi}(e^\beta+c_\theta)}
\left(\begin{bmatrix}
1\\
-1
\end{bmatrix}\otimes \mat{I}+\begin{bmatrix}
1\\
1
\end{bmatrix}\otimes \mat{Y}\right)
,\qquad\vec{\mu}=\begin{bmatrix}
\mu_1\\
\mu_2
\end{bmatrix}.
\end{align}
The integral can now be solved using formula Eq.~\eqref{eq:multivari_gaussian_int}:
\begin{align}
\int_{-\infty}^\infty\frac{d^2\mu}{\pi}e^{-\frac{1}{2}\vec{\mu}^\tp(2\mat{I})\vec{\mu}}\Theta\begin{bmatrix}
-(\mat{I}\otimes\mat{\Omega})\vec{u}_{\paulivec\paulivec'}^{+}\\
\vec{u}_{\paulivec\paulivec'}^{-}
\end{bmatrix}\left(\mat{V}\vec{\mu};\mat{I}\otimes\mat{\tau}^\het\right)=\Theta\begin{bmatrix}
-(\mat{I}\otimes\mat{\Omega})\vec{u}_{\paulivec\paulivec'}^{+}\\
\vec{u}_{\paulivec\paulivec'}^{-}
\end{bmatrix}\left( \vec 0;\mat{I}\otimes\mat{\tau}^\het+\pi i\mat{V}\mat{V}^\tp \right).
\end{align}
With this, the process matrix is given by
\begin{align}
    \chi_{\paulivec\paulivec'}(\vec{m}) 
    &= 
     \frac{1}{4} 
    \frac{e^{2\beta} e^{\pi i(\vec{u}^{-}_{\paulivec\paulivec'})^\tp(\mat{I} \otimes \mat{\Omega}) \vec{u}_{\paulivec\paulivec'}^{+}} }{ \pi (c_\theta + e^\beta )^2}
    \Theta\begin{bmatrix}
    -(\mat{I}\otimes\mat{\Omega})\vec{u}_{\paulivec\paulivec'}^{+}\\
    \vec{u}_{\paulivec\paulivec'}^{-}
    \end{bmatrix} \left(\vec 0;\mat{T}\right),
\end{align}
where
\begin{align}
\mat{T} &\coloneqq \mat{I}\otimes \mat{\tau}^\het - \frac {s^2_{\theta} }{(c_\theta + e^\beta)^2} (i\mat{X} \otimes \mat{I} + \mat{\Omega} \otimes \mat{\Omega}).
\end{align}

\twocolumngrid

%

%
%
%
%
%
%
%
%
%
%
%
%
%
%
%
%

%

%
%
%
%
%
%
%
%
%
%
%
%
%

\end{document}